\documentclass[10pt]{iopart}

\usepackage{graphicx, subcaption}
\usepackage{amssymb}
\usepackage{xcolor, ulem} 

\newcommand{\add}[1]{#1}
\newcommand{\del}[1]{}
\newcommand{\adda}[1]{#1}
\newcommand{\dela}[1]{}

\begin{document}

\title[Dynamics of NRP Discharge driven by flame passage]{Nanosecond Pulsed Discharge Dynamics During Passage of a Transient Laminar Flame}

\author{Colin A. Pavan and Carmen Guerra-Garcia}

\address{Massachusetts Institute of Technology, 77 Massachusetts Ave, Cambridge, MA, USA}
\ead{cpavan@mit.edu}
\vspace{10pt}

\begin{abstract}
This work presents an experimental study of a nanosecond repetitively pulsed dielectric barrier discharge interacting with a transient laminar flame propagating in a channel of height near the quenching distance of the flame. The discharge and the flame are of comparable size, and the discharge is favoured at a location where it is coupled with the reaction zone and burnt gas. The primary goal is to determine how the discharge evolves on the time scale of the flame passage, with the evolution driven by the changing gas state produced by the moving flame front. This work complements the large body of work investigating the effect of plasma to modify flame dynamics, by considering the other side of the interaction (how the discharge is modified by the flame). The hot gas produced by the combustion had a strong effect on the discharge, with the discharge preferentially forming in the region of hot combustion products. The  per-pulse energy deposited by the discharge was measured and found to increase with the size of the discharge region and applied voltage. The pulse repetition frequency did not have a direct impact on the per-pulse energy, but did have an effect on the morphology and size of the discharge region. Two distinct discharge regimes were observed: uniform and filamentary (microdischarges). Higher pulse repetition frequencies and faster-cooling combustion products were more likely to transition to the filamentary regime, while lower frequencies and slower-cooling combustion products maintained a uniform regime for the entirety of the time the discharge was active. This regime transition was influenced by the ratio of the time scale of fluid motion to the pulse repetition rate (with no noticeable impact caused by the reduced electric field), with the filamentary regime \adda{preferentially} observed in situations where this ratio was small. This work demonstrates the importance of considering how the discharge properties will change due to combustion processes in applications utilizing plasma assistance for transient combustion systems.
\end{abstract}

%
\vspace{2pc}

\vspace{2pc}
Submitted to \PSST
%
%
\ioptwocol

\section{Introduction and Motivation}

Recent years have seen considerable interest in plasma assisted combustion (PAC) technologies. The plasma interacts with the combustion processes in multiple ways, and has shown particular promise in enabling or stabilizing combustion in difficult regimes near flammability limits. These include lean flammability limit extension and stabilization of lean flames \cite{Lacoste2013, Blanchard2021, Starikovskaia2006a, Pilla2006, Barbosa2015} and low residence time high-speed or supersonic combustion \cite{Ju2015,Starikovskiy2013,Do2010, Do2010a}. In the related field of plasma assisted ignition (PAI), the application of plasma has been shown to reduce ignition delay times \cite{Aleksandrov2009, Castela2016, Cathey2007, Kosarev2008} and enhance low-temperature chemistry \cite{Sun2014, Yang2017}. The plasma acts in many different ways depending on the way it is generated, the neutral gas composition and state, and the magnitude of the reduced electric field, E/N. Lower reduced electric fields, below around $10^2$Td in air plasmas \cite{Nagaraja2013}, will put energy into rotational and vibrational excitation modes, while higher fields, on the order of $10^2$-$10^3$Td, can cause electronic excitation,  electron-impact dissociation and ionization \cite{Popov2016, Starikovskiy2015}. \add{Experimentally determining the reduced electric field can be challenging, but recent advances using the electric field induced second harmonic generation (E-FISH) technique have allowed for direct measurement of this quantity \cite{Orr2020, Rousso2020}.}

These different mechanisms of interaction, and the modification of the combustion process as influenced by the plasma, are also strongly dependent on the positioning of the plasma with respect to the flame front. Several works have focused on the influence of plasma generated upstream of the flame front, in the reactant stream. Stange et al. \cite{Stange2005} observed that an AC dielectric barrier discharge (DBD) in the microdischarge mode applied to the fuel flow in a coaxial air-propane nonpremixed flame configuration would cause a flame speed increase, but the authors did not provide details into the potential mechanisms enabling the interaction. More recent work by Elkholy et al. \cite{Elkholy2018} and Evans et al. \cite{Evans2017} apply nanosecond pulsed plasma upstream of lean, premixed methane air flames, with the former using a DBD configuration and the latter a pin-to-pin configuration, and also observe flame speed increase. Typical per-pulse energies used were \del{$\mathcal{O}(100\mu J)$} \add{of order $10^{-4}$J} \del{and the}\add{with} pulse repetition frequencies up to 10kHz for actuation on flames with power \del{$\mathcal{O}(100W)$} \add{of order $10^2$W}. Other works have investigated plasma generated in-situ with the flame (e.g. overlapping with the flame reaction front). This environment is considerably more complex; not only because the plasma chemistry will more directly interact with the combustion chemistry, but also because the presence of the flame front will have an effect on the discharge properties through the reduction in gas density and corresponding increase in E/N, \add{and due to changing ionization properties linked to combustion chemistry \cite{Adamovich2009, Zhong2019, Rousso2020}}. From the plasma perspective, there has been some work investigating these environments; for example \cite{Guerra-Garcia2014,Guerra-Garcia2015} looked at how discharges would preferentially form in regions of spatially inhomogeneous gas that favoured breakdown. Other works \cite{Li2013, Nagaraja2015, Massa2017} have looked at how discharge energy is deposited primarily in regions heated by the gas combustion.

This work will focus on plasma generated using the nanosecond repetitively pulsed dielectric barrier discharge (NRP DBD) method. This strategy is attractive for plasma assisted combustion applications because it can generate high reduced fields (several hundred Townsend) at a low average power \cite{Adamovich2009a, Adamovich2009, Nikandrov2008, Lou2007}. NRP DBD is also capable of producing a diffuse, uniform discharge at atmospheric pressures, which is advantageous in many applications. However, there are many factors, particularly in a combustion environment, that can cause a transition to a filamentary microdischarge structure. One of these factors is the strength of the electric field; Liu et al. give a threshold of 130kV/cm for room temperature air at atmospheric conditions \cite{Liu2018, Liu2019}, equivalent to 530 Td, for single nanosecond DBD pulses. The composition of the mixture will also play a role in the uniformity, with the presence of hydrocarbons and NO having been observed to more readily cause a uniform to filamentary transition \cite{Adamovich2009, Rousso2020}. A proposed mechanism for this transition is a coupled plasma thermal-chemical instability \cite{Zhong2019, Rousso2020, Zhong2021}. Other contributing effects include the gas residence time in the discharge region, with longer residence times allowing the microdischarge instabilities to more readily develop \cite{Khomich2016, Zhang2021}. \add{This may be related to residual space and surface charge from previous discharges creating a memory effect in the discharge region \cite{Huang2014,Huang2018,Huang2020}.}

Changes in discharge structure driven by gas condition are particularly relevant in situations involving transient combustion phenomena and moving flame fronts where in-situ generated plasma for PAC may be of interest. Examples include internal combustion engines, rotating detonation engines and static or dynamic flame stability studies \add{\cite{Moeck2013,Kim2015,Zhu2017, Cathey2007,Shanbhogue2022}}. While there is considerable work looking at how plasma affects combustion processes, the other side of the interaction, namely how the discharge is affected by the transient combustion environment, has been studied considerably less. There is a significant reduction in the neutral gas number density as a flame front passes, and this creates a corresponding increase in the reduced electric field. This change in reduced field, on the order of 4-7 times (for a gas starting at room temperature heated to 1200-2100K), will dramatically affect the discharge behaviour. More subtly, the flame passage will also modify the gas composition and flow field, which will also influence the discharge behaviour. In cases where transient flame fronts are expected or situations looking at flame stabilization where the flame location may change with time, it is important to consider how the change in flame position and operating conditions will modify the discharge behaviour to effectively design a control scheme for plasma assistance that accounts for the evolving environment. This work looks at the evolution of a coupled flame-discharge system as a flame front propagates in between two dielectric-covered electrodes in a narrow channel to investigate how the discharge is affected by the transient environment created by the passing flame. Specific attention is given to how the discharge evolves on the time scales of the flame propagation.

\section{Setup and Methodology}

\subsection{Experimental Setup}

\begin{figure*}
    \centering
    \includegraphics[width=\textwidth,trim={0cm 6.5cm 0cm 7cm}, clip]{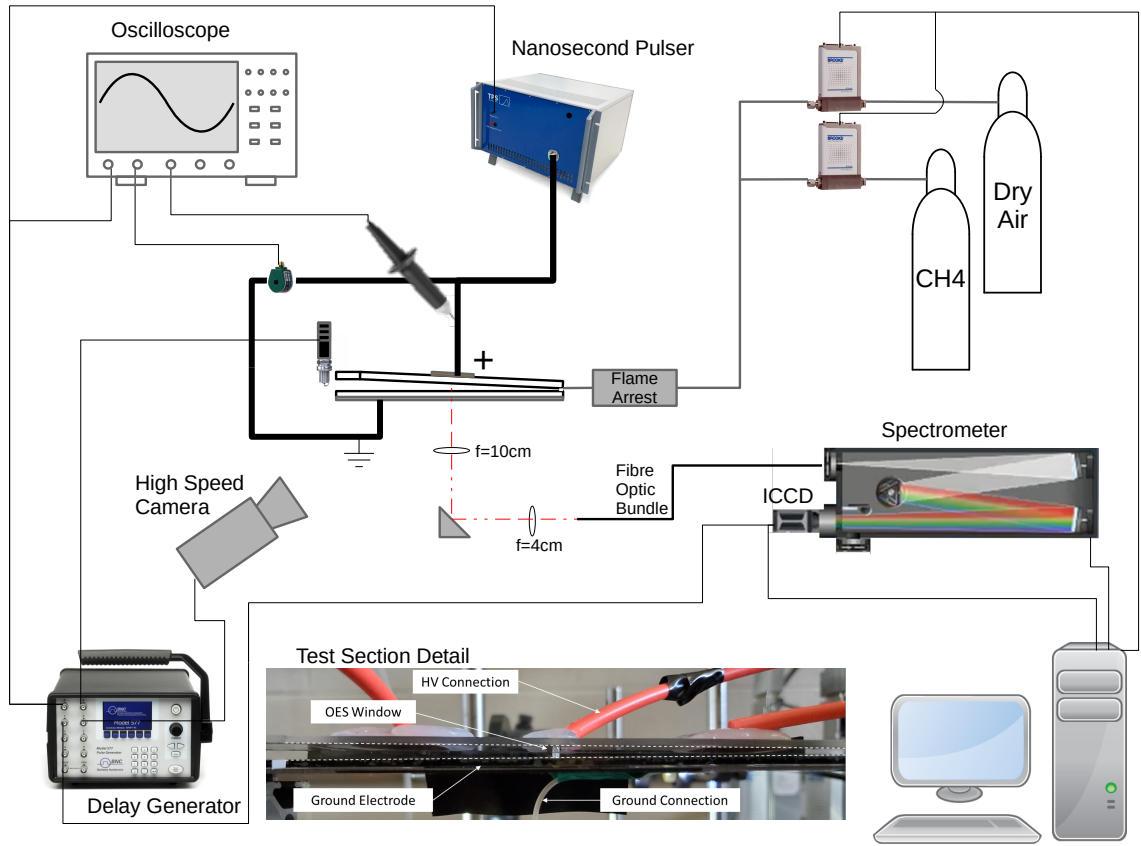}
    \caption{Schematic of the experimental setup with image of test section shown in inset.}
    \label{fig:exp_schematic}
\end{figure*}

The experimental setup is shown in figure~\ref{fig:exp_schematic}. The main test section is a tapered quartz channel with a width of 3cm and a height varying from 4.2mm to 1mm over 20cm. Premixed methane and air enters from the small end of the taper and flows towards the large end. At the exit of the test section, a conventional spark plug is used to ignite the mixture. The flame then flashes back through the channel and is thermally quenched by the narrowing walls of the chamber. This geometry was designed for experiments on quenching distance modification using plasma assistance \del{hence the taper,} \add{where the taper would allow the discharge to be applied at heights varying continually from above, at or below the quenching height of the flame, similar to the configuration of \cite{Murphy2014}. The aspect ratio greater than ten is sufficient to minimize the effect of the channel width on quenching height \cite{Berlad1957}.} \del{but} The height change is very gradual and over short distances the opposite faces can be considered approximately parallel (the height changes by less than 0.2mm over 1cm). \add{Having the flame propagate counter to the flow of gas slowed its propagation in the lab-frame which was advantageous for the time-resolved measurements.} The construction out of quartz allows optical access and also acts as a dielectric barrier for a DBD. The gas mixture flowing into the test section is controlled by two Brooks GF40 mass flow controllers. High voltage is applied using a nanosecond pulse generator (Transient Plasma Systems SSPG-20X-HP1). The high voltage electrode is made of copper tape attached to the upper 2mm thick quartz plate. The ground electrode is an aluminum plate that also acts as a support for the test section. The size and position of the top electrode could be varied; results in this work are reported for two cases. The \emph{long electrode} is 30mm in length stretching over a region of the test section where the height changes from 3.4 to 2.9mm, and the \emph{short electrode} is 5mm in length, centered over the point where the channel height is 3.1mm. Note that thermal quenching of the flame for a stoichiometric mixture happens at around 2.3 mm of height for this experiment configuration. The free parameters in any set of tests are the pulse repetition frequency (f), voltage (V), number of pulses ($N_p$), equivalence ratio ($\mathrm{\phi}$) and total flow rate (Q).

The system has three independent diagnostics systems, which are described in detail in the following section. A high speed camera (Edgertronic SC2+) is used for visual inspection of the flame and discharge. Electrical measurements are taken using a high voltage probe (Lecroy PPE 20kV\adda{, 100MHz BW}\footnote{\adda{BW is the -3dB high frequency bandwidth reported in the equipment specifications.}}) and a Rogowski-coil style current monitor (Pearson model 2877\adda{, 200MHz BW}) placed on the current return side of the circuit. Signals are collected at 100ps resolution using a Teledyne Lecroy Waverunner 9254 oscilloscope \adda{(500MHz BW). Dominant frequency components in current and waveform signals were $\leq$50MHz}. The discharge is also observed with an optical emission spectroscopy system. The light collection is done using two planar-convex lenses which focus the discharge emission from a 0.8mm long region of the channel onto a fiber optic bundle that is attached to the spectrometer (Acton SpecraPro 2750). The spectra are recorded using a Princeton Instruments PIMAX4 intensified CCD camera.

The experiment is operated by continuously flowing gas at the desired flow rate and equivalence ratio. Since the flame propagates opposite the direction of gas flow, the lab-frame flame speed is slower than the laminar flame speed. The gentle taper of the channel and low flow rates ($\leq$1000sccm\footnote{\emph{standard conditions} are referenced to 21.1$^o$C and 1013.25mbar per the flow controller manufacturer.}) ensure that the unburnt gas flow remains laminar and slower than the laminar flame speed at all points along the taper so that the flame will continue moving upstream. \add{Signal synchronization is done using a delay generator (Berkeley Nucleonics Corp. Model 577-4C). A test begins by simultaneously triggering the spark plug to ignite the mixture and starting the high speed video. A train of pulses (typically 1000-}\adda{4000} \dela{2000} \add{long) is initiated after some delay, with the ICCD gate synchronized to capture the pulse. All pulse timing can then be referenced to the start time of the high speed video.}

\subsection{Diagnostics}
\subsubsection{Visual Flame Tracking}
\begin{figure}[]
    \centering
    \begin{subfigure}[]{0.45\textwidth}
        \centering
        \includegraphics[width=\textwidth]{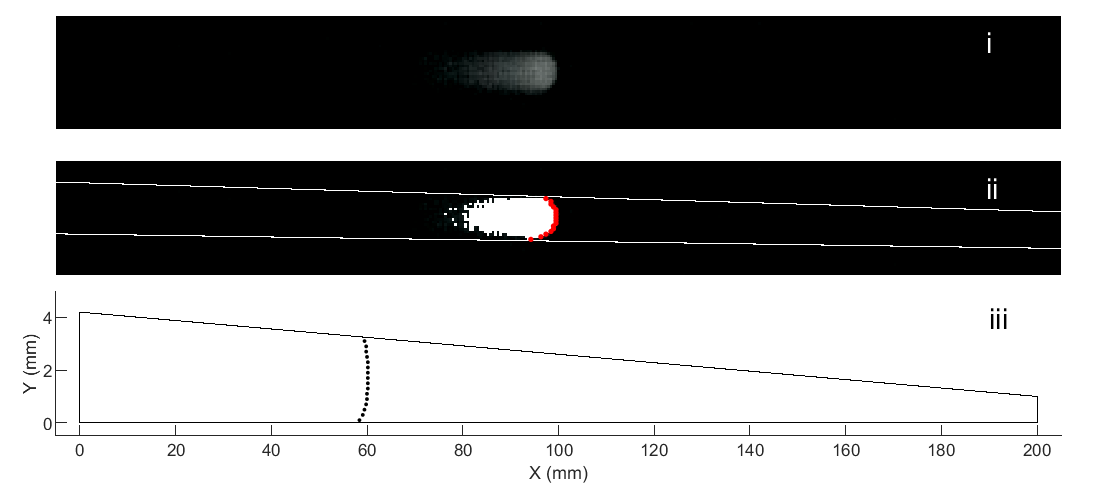}
        \caption{Flame Tracking Algorithm. (i) Raw video, (ii) Flame identified in channel, flame front marked in red, (iii) Flame front mapped to channel dimensions.}
        \label{fig:flame_tracking}
    \end{subfigure}
    \begin{subfigure}[]{0.45\textwidth}
        \centering
        \includegraphics[width=\textwidth]{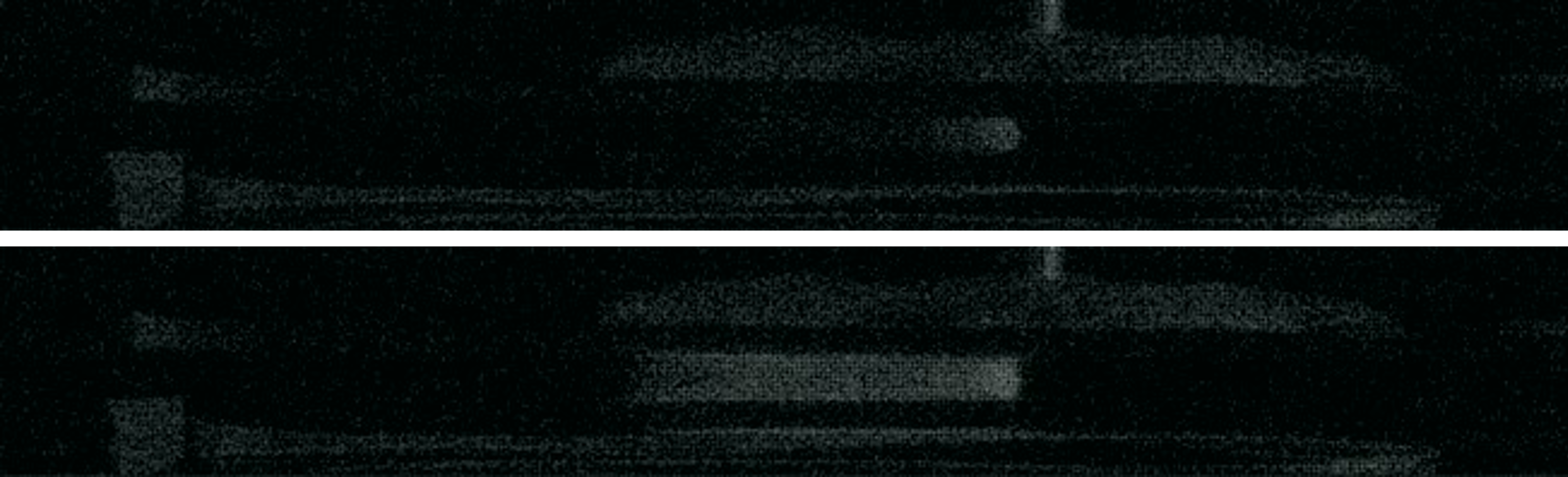}
        \caption{Two sequential frames from a video with a frame rate twice the pulse repetition rate showing alternate no discharge (top) and discharge (bottom) frames.}
        \label{fig:HFR_demo}
    \end{subfigure}
    \caption{Analyses performed using the high speed video.}
    \label{fig:videography}
\end{figure}

Flame propagation and discharge structure are observed using high speed videography. The frame rates used in this experiment can resolve events that occur on the timescale of the flame propagation \del{($\mathcal{O}(1 $ms))}\add{(milliseconds)} and on the timescale of pulse repetition \del{($\mathcal{O}(100\mathrm{\mu}$s))}\add{(hundreds of microseconds)}. Fiducial markers were placed on the test section and the camera was focused to give a spatial resolution of approximately 0.2mm$\times$0.2mm per pixel. With a lab-frame flame speed of $\leq 40$cm/s, a camera frame rate of 2000 frames per second (2kfps) ensures that the flame will travel at most one pixel per frame, and resolution in flame front location is limited by the camera pixel density and not the frame rate. To visually resolve individual pulses, a higher frame rate is needed. The pulse repetition frequency used in this work was between 2-10kHz. \del{The duration of an individual pulse is $\mathcal{O}(10 $ns)}. By operating the camera at a frame rate twice the pulse repetition rate (referred to in this work as the high frame rate, or HFR, mode), individual pulses could be observed with alternating frames with and without a discharge. The camera exposure is approximately the inverse of the frame rate (orders of magnitude longer than the pulse duration), so the images with a discharge are the integrated emission of the discharge over the entire event. Operating in this mode was useful for separating the flame from the discharge in the monochrome images, and for discerning how sequential pulses differ (for example, in the location of filamentary microchannels). Figure~\ref{fig:videography} shows the analyses performed using the high speed video. \adda{The intensity gain of the camera was adjusted based on the frame rate and additional brightening was performed in post-processing to ensure the relevant structures were clearly visible; all videos taken at 2kfps used an ISO gain of 25000 with an addition 1.5x artificial brightening in post processing. This includes all flame images shown in this work, with the exception of figure~\ref{fig:videography}.}

\subsubsection{Energy Measurements}
\add{
The energy was calculated by the product of the voltage and current integrated over the initial pulse and all subsequent reflections according to equation~\ref{eqn:nrg_calc}. 
\begin{equation}\label{eqn:nrg_calc}
    W_{\mathrm{p}}=\int_{t_0}^t V(t^\prime)I(t^\prime-t_{\mathrm{lag}})\mathrm{d}t^\prime
\end{equation}
$t_{lag}$ accounts for the signal travel time differences, which must be considered to get accurate energy measurements on the nanosecond timescale, and $t_0$ is chosen just before the initial voltage rise. The reflections are caused by a mismatched impedance; the pulser is rated for a 50$\Omega$ load but the variable impedance of the discharge gap (due to the changing gas condition and presence or lack of residual electrons from the previous pulse) mean this condition is rarely met. These reflections are common, with works such as \cite{Pancheshnyi2006} displaying a similar behaviour. Technical details on how the energy measurements were performed, including the method for determining $t_{lag}$, are given in \ref{sec:app_energy}.
}
\subsubsection{Optical Emission Spectroscopy}\label{sec:OES_methods}

Optical emission spectroscopy (OES) is performed to probe the temperature of the coupled flame and discharge.  \adda{Temperature measurements are used for two main purposes in this work. The first is to determine if any gas heating is caused by the discharge and the second is to estimate the reduced electric field within the discharge gap, which is a key parameter governing the discharge.} 

Due to fast relaxation times (\del{$\mathcal{O}(1ns)$}\add{of order 1ns} \cite{Rusterholtz2013}), and comparatively long shutter gate width ($1\mu s$), rotational-translation (RT) equilibrium is assumed. The RT temperature is determined from the N$_2$ C-B (0,2) transition around 380nm \cite{Laux2003}. An 1800 groove/mm grating blazed at 500nm and centered on 376.5nm was used for these measurements. \del{Vibrational temperatures were determined by fitting multiple N$_2$ emission peaks between 347nm and 402nm using a 300 groove/mm, 500nm blaze grating centered at 375nm. Looking at a large number of peaks in this range gives better temperature resolution than looking at only the relative intensity of the two peaks that were visible with the 1800 groove/mm grating \cite{Rusterholtz2013}. For this measurement, a relative intensity calibration had to be performed since over the 55nm range observed the grating and ICCD camera had a non-negligible change in sensitivity to the different wavelengths. The calibration was performed with an Oriel Instruments 63355 quartz tungsten halogen lamp.}

The ICCD camera shutter is opened for 1$\mathrm{\mu}$s for each pulse and several sequential spectra are acquired on the sensor before readout. Light was collected from a window 0.8mm wide on the test section centered about the point where the channel height was 3.1mm, and the number of accumulations was set to integrate the light from pulses occurring within a 6ms time period. \dela{Due to the transient passage of the flame, at a rate of approximately 20cm/s, this meant that the spectra collected}\adda{At this location, the flame was moving at approximately 20cm/s in the lab frame, which meant it would travel 1.2mm during the time it took to accumulate the light for one spectra. Therefore, measured relative to the position of the flame front, the collected spectra} are the average for a region 2mm wide in the flame propagation direction (window size + distance flame moves). Spectra were acquired every 12.5ms, which was the fastest rate possible with the equipment used (6ms acquisition, 6.5ms readout).The finite dimension of the channel and Taylor-Saffman instabilities \cite{Saffman1958, Murphy2015} meant that the flame was also not a perfect planar front, so there is some additional non-uniformity caused by\del{the curved flame front} \add{instabilities in the camera line of sight direction. \add{In addition, the flame has a crescent shape, as seen in figure~\ref{fig:flame_tracking} (ii), which is typical of flames in laminar flow in small dimension channels \cite{Lewis1987, Ju2006}.} The temperature rise caused by a single discharge pulse in a NRP DBD should be on the order of a few Kelvin at most \cite{Lefkowitz2015, Rousso2020}, so the $1\mu s$ gate window and  integration over several pulses introduces minimal error to RT temperature measurement}.  

\begin{figure}[]
    \centering
    \includegraphics[width=0.48\textwidth, trim={1cm 1.5cm 1cm 1cm}, clip]{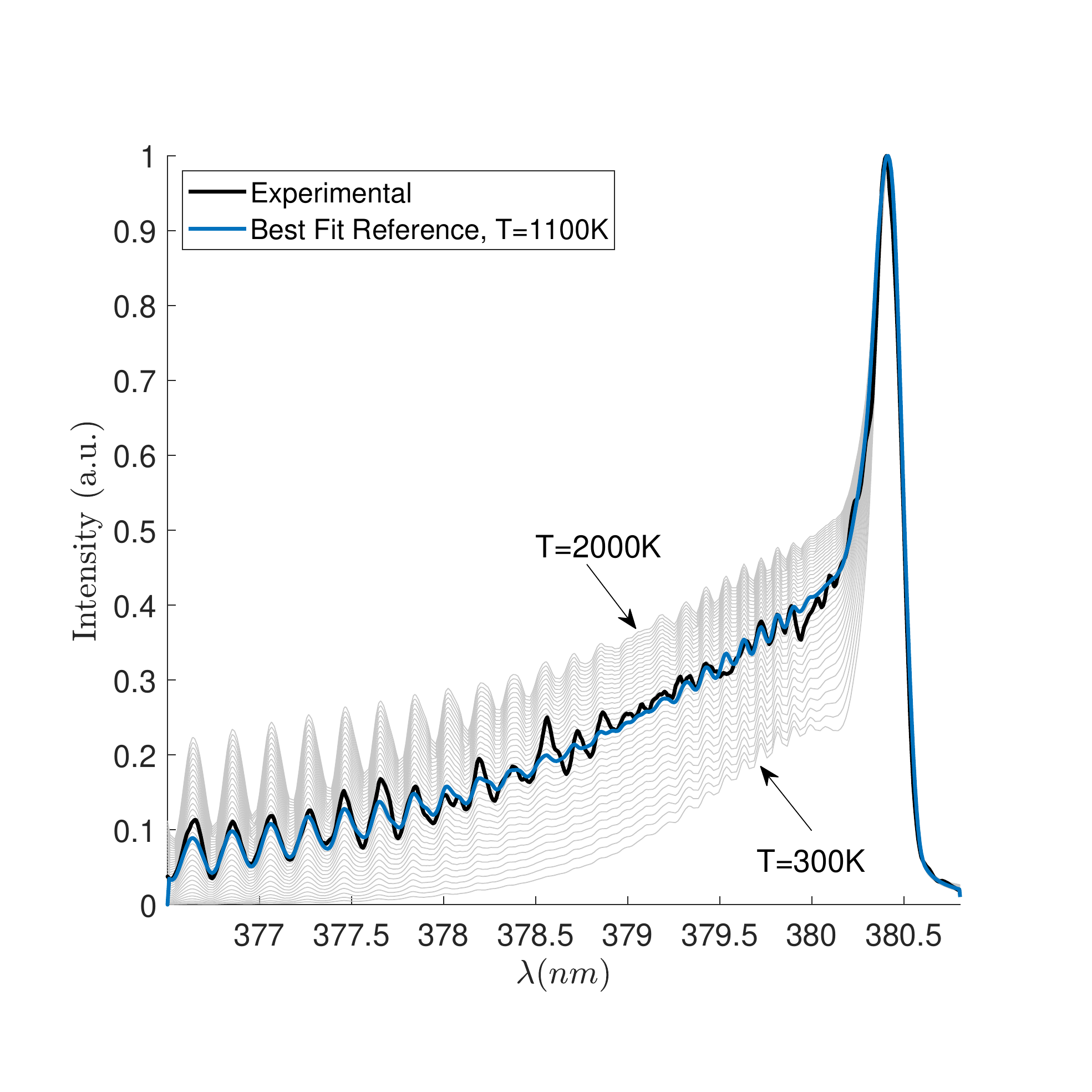}
    \caption{Example of spectral fitting procedure. Experimental spectrum is compared to each of the synthetic spectra (gray lines) generated at 50K increments using SPECAIR.}
    \label{fig:spectral_fitting}
\end{figure}

Spectral fitting is done using SPECAIR \cite{Laux2003}. Each test generates 20 spectra and dozens of tests are performed; this creates thousands of spectra. To automate the OES RT temperature measurements, synthetic spectra were generated at 50K increments, which is approximately the resolution in RT measurement that can be obtained by this method \cite{Laux2003}. The synthetic spectra were generated for wavelengths in the range 376.5-380.7nm, a region insensitive to vibrational temperature.  Each of these synthetic spectra were then compared with the experimental spectra and the best fit was chosen based on finding the temperature that maximized the cross correlation between the (normalized) signals. An example of the procedure for determining RT temperature by spectral fitting is shown in figure~\ref{fig:spectral_fitting}. \add{Assuming that the fitting procedure determines the best fit to within 50K and an additional 50K uncertainty inherent in determining the R-T temperature from fitting the C-B (0,2) transition using SPECAIR \cite{Laux2003}, the R-T temperature can be taken as accurate to within $\pm100$K}.\del{This procedure determines the temperature to within approximately 50K}.

\section{Dynamic Evolution of a \dela{RPN}\adda{NRP} DBD During Flame Passage}
The experiments in this section were obtained using the \emph{long electrode}.
\subsection{Evolution of the Discharge Appearance}
\begin{figure}
    \centering
    \includegraphics[width=0.48\textwidth, trim={1cm 0 0.5cm 0},clip]{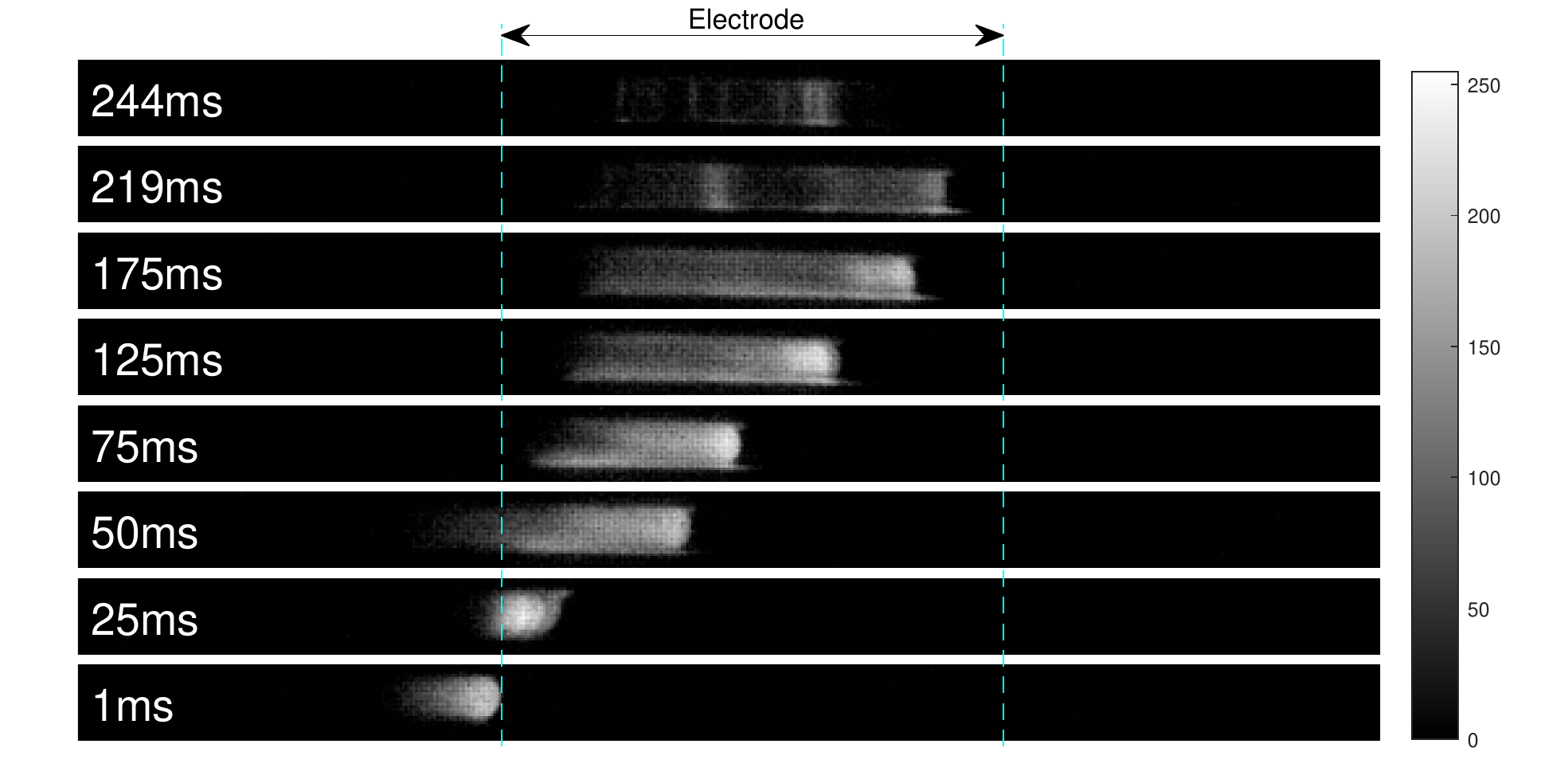}
    \caption{Evolution of the discharge appearance during flame passage. Video was taken using long electrode at a frame rate of 2kfps. Flame moves left to right and channel height decreases from left to right. Time 0ms corresponds to the application of the first voltage pulse  \adda{and pulsed voltage is applied during all frames. Conditions: f=8kHz, V=13.8kV, $\mathrm{\phi}$=1, Q=1000sccm, $N_p$=2000. Legend in colorbar is in arbitrary units.}}
    \label{fig:long_elec_evolution}
\end{figure}

Using the long electrode, it is possible to observe the discharge evolution in both space and time as the flame passes underneath it. This is shown in figure~\ref{fig:long_elec_evolution} \adda{ for an example case}. \dela{The example is given for the case of pulses at f=8kHz, V=13.8kV and flow conditions of $\mathrm{\phi}$=1, Q=1000sccm.} When the high voltage pulses begin, there is no visible discharge in the inter-electrode gap and the only luminosity comes from the flame (frame at 1ms). By 25ms, the flame has entered the discharge region and the shape begins to change. The initially curved leading edge flattens out. Using the HFR visualization, it was observed that the flame itself maintains its curved structure and the apparent flattening of the front is caused by the superimposed discharge. As time progresses, the size of the luminous region increases. This is caused by a uniform discharge in the burnt gas behind the flame. The uniform discharge extends from the flame front itself to the edge of the electrode, see the 50ms frame. By 75ms, the trailing end of the discharge has started to separate from the edge of the electrode. There are two reasons for this. First, the taper of the channel means that the gap at the left end of the electrode is slightly larger than at the right end, so the electric field will be slightly lower. Second, the further behind the flame front, the more the gas will have cooled against the walls of the chamber. This cooling causes the density to increase and the reduced electric field to decrease. These combined effects cause the discharge to form preferentially closer to the flame front. This trend continues for the next 100ms, with the discharge appearing stronger closer to the flame front. Note that the region of increased luminosity at the very right end of the discharge region is the flame itself (this was confirmed by the HFR videos). Up until this point, the entire discharge remains uniform.

Between 175ms and 219ms the flame reaches its \dela{maximum extent}\adda{furthest most position} and is quenched before it exits the inter-electrode region. \adda{This quenching occurs when the channel height is greater than the typical quenching distance of a flame at the given equivalence ratio, and is caused by the discharge itself. Further discussion of this premature quenching caused by the discharge in this setup can be found in~\cite{Pavan2021a}.} As soon as quenching occurs, the heat release quickly drops and the gas rapidly cools. A regime transition is observed, transitioning from the uniform discharge that was present throughout the burnt gas to filamentary structures. At 219ms, filaments are beginning to form but there is still a relatively uniform regime near where the flame front was. By 244ms, the uniform regime has almost completely disappeared and the discharge gap is dominated by the filamentary regime. At this point, the discharge is no longer preferentially forming at the right end of the electrode, where the flame front had previously been and where the height is lowest. Instead, it occurs closer to the center of the electrode. The cause of this regime transition is discussed further in section~\ref{sec:regime_transition}.

\subsection{Evolution of the Per-Pulse Energy}

\begin{figure}[]
    \centering
    \includegraphics[width=0.45\textwidth, trim={1.3cm 0 1.5cm 0},clip]{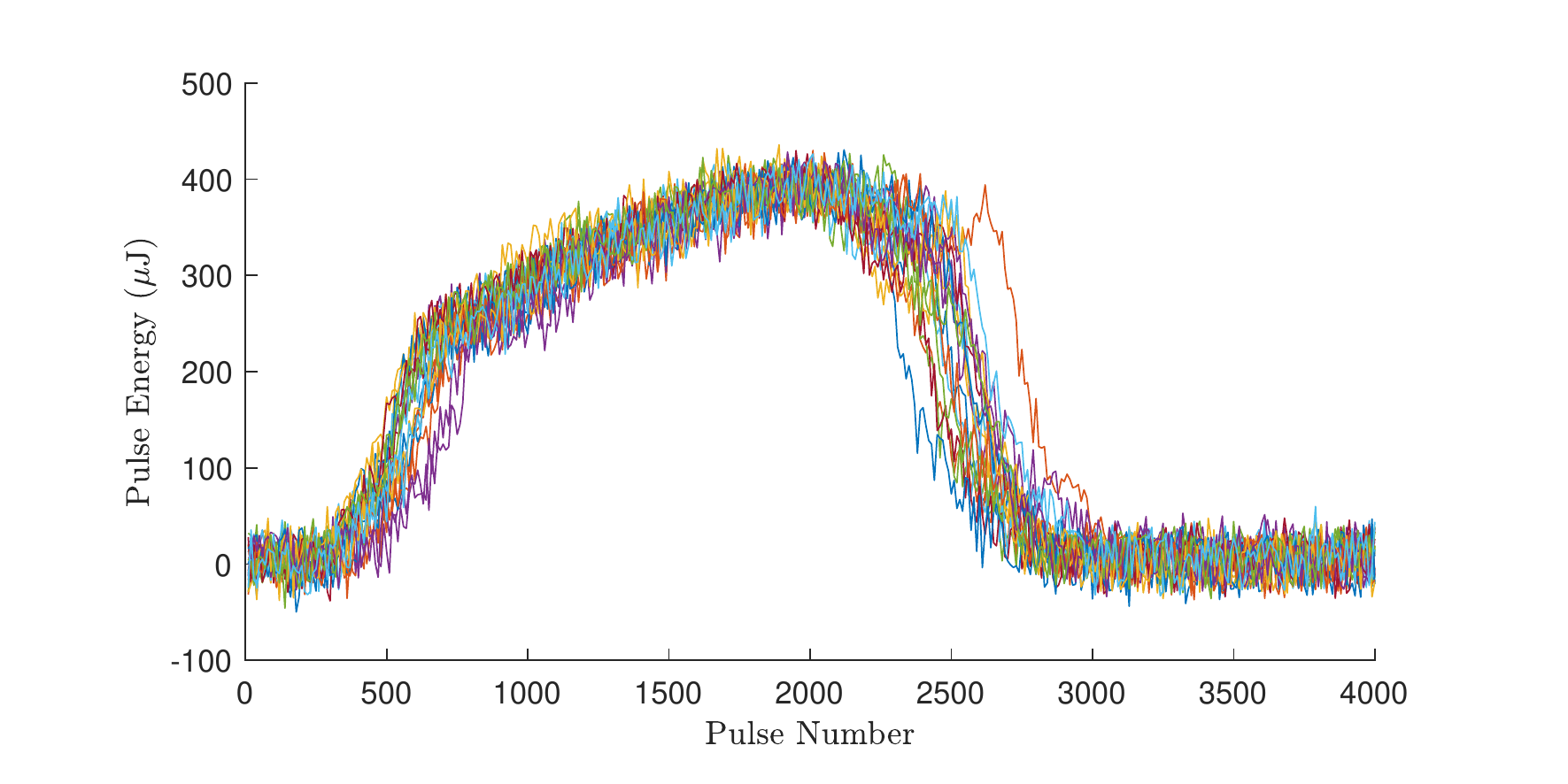}
    \caption{Evolution of the per-pulse energy using the long electrode. f=10kHz, V=12.9kV,  $\mathrm{\phi}$=1, Q=1000sccm, \adda{N$_p$ = 4000}.}
    \label{fig:long_elec_nrg}
\end{figure}

The energy deposited by the discharge as a function of pulse number is shown in figure~\ref{fig:long_elec_nrg}, for \add{an example case.} \del{two different configurations in the electrical parameters and similar flow conditions.} Each coloured line corresponds to an individual test in a sequence of 20 independent tests. The flame enters the inter-electrode region around pulse 400. \del{in both cases. Both cases show similar behaviour in the early stage of the discharge evolution.} There is an initial region of rapid energy increase caused by the discharge growing in size. Around pulse 700 \del{in both cases} there is an inflection point. This corresponds to the discharge continuing to grow as the flame front penetrates further into the inter-electrode region, but also beginning to shrink at the wider end of the taper as the cooling gas and larger gap makes it  more difficult to sustain the discharge (see discussion in preceding section). The net effect is still an overall increase in the size of the discharge, but at a slower rate than previously. Eventually, a peak is reached. In the case \del{of 10kHz and 12.9kV,} \add{shown in} figure~\ref{fig:long_elec_nrg}, this corresponds to the flame quenching. After this, as there is no longer a continuous heating source and the gas cools, there is a rapid decay in the energy per-pulse and the discharge transitions first to a filamentary structure and ultimately disappears because the applied voltage is insufficient to maintain a discharge in cold gas. \adda{When a lower-power discharge was applied, the flame would pass through the electrode, although would be weakened by the discharge. In these cases, there would be a decrease in per-pulse energy as the flame weakened followed by an increase in per-pulse energy once the flame had passed through the electrode. A more thorough discussion of this phenomenon is given in \ref{sec:app_reinvig}.} \dela{In other cases, the flame passed through the electrode although the discharge tended to weaken it. Details of one such case are given in \ref{sec:app_reinvig}.} \del{the case of 6kHz and 11.7kV, figure~\ref{fig:long_elec_nrg_a},  the flame becomes very weak as it travels along the electrodes and this causes the discharge region to shrink. After the flame has left the electrode, around pulse 1700, the flame strengthens again and this causes a re-ignition of the discharge which accounts for the increase in per-pulse energy. This effect is described in more detail in the following section.}

\del{\bf{3.3 Discharge Re-ignition Driven by Invigoration of Flame}}

\section{Local Measurements to Reveal Plasma-Flame Interaction}
The experiments in this section were obtained using the \emph{short electrode}.
\subsection{\dela{Visual Evolution of Discharge}\adda{Advantages of Short Electrode}}
The preceding experiments looked at the evolution of the discharge throughout the burnt gas region \adda{and with the discharge remaining coupled to the flame front. Those experiments show how the coupled flame-discharge system evolve simultaneously in both space and time. The length scale of spatial variation is much smaller than the length of the electrode, so it is possible to observe how the discharge behaves differently in front of, on top of, and at various distances behind the flame at the same time. This configuration was well-suited to performing visual analysis of the discharge evolution using the high speed video. However, the spatial inhomogeneity makes it difficult to perform more quantitative diagnostics like discharge energy measurements. In particular, when there are both filamentary and uniform regimes present or when the uniform region has a noticeable gradient in luminosity, it is unknown how the energy deposition varies along the electrode and with the changes in discharge structure. To better understand the spatial variations in the system, the short electrode was used. With the 5mm electrode, } \dela{The finite time it took for the flame to pass under the electrode played a role in the evolution of the discharge. For this section, the short electrode is used. The 5mm length of the electrode meant that the time required for the flame to traverse the electrode was much shorter and} the discharge evolution is primarily governed by the changing gas condition underneath the electrode as the hot gas behind the flame cools. Except for the time when the flame front passes through this region, the short electrode length ensures quite uniform background gas, \dela{so temperature measurements are meaningful}\adda{which meant that when discharge energy and temperature measurements were taken they could be directly associated with a specific visual appearance and location}. The other significant effect here is that the flame would pass under the electrode and continue down the tapered channel in all cases, and the high voltage would shut off before the flame had quenched. This was opposite what happened with the long electrode; in that case the application of the high voltage often caused the flame to quench prematurely, before it exited the electrode. This results in a different gas cooling behaviour. In the case of the long electrode, the hot gas cooled against the walls of the chamber and at the same time the thermal source (the flame reaction front) disappeared, causing rapid cooling. In the case of the short electrode, only the former effect is present so the gas cooling occurs more slowly.

\subsection{\adda{Visual Evolution of Discharge}}

\begin{figure}
    \centering
    \includegraphics[width=0.48\textwidth, trim={1cm 0 1cm 0},clip]{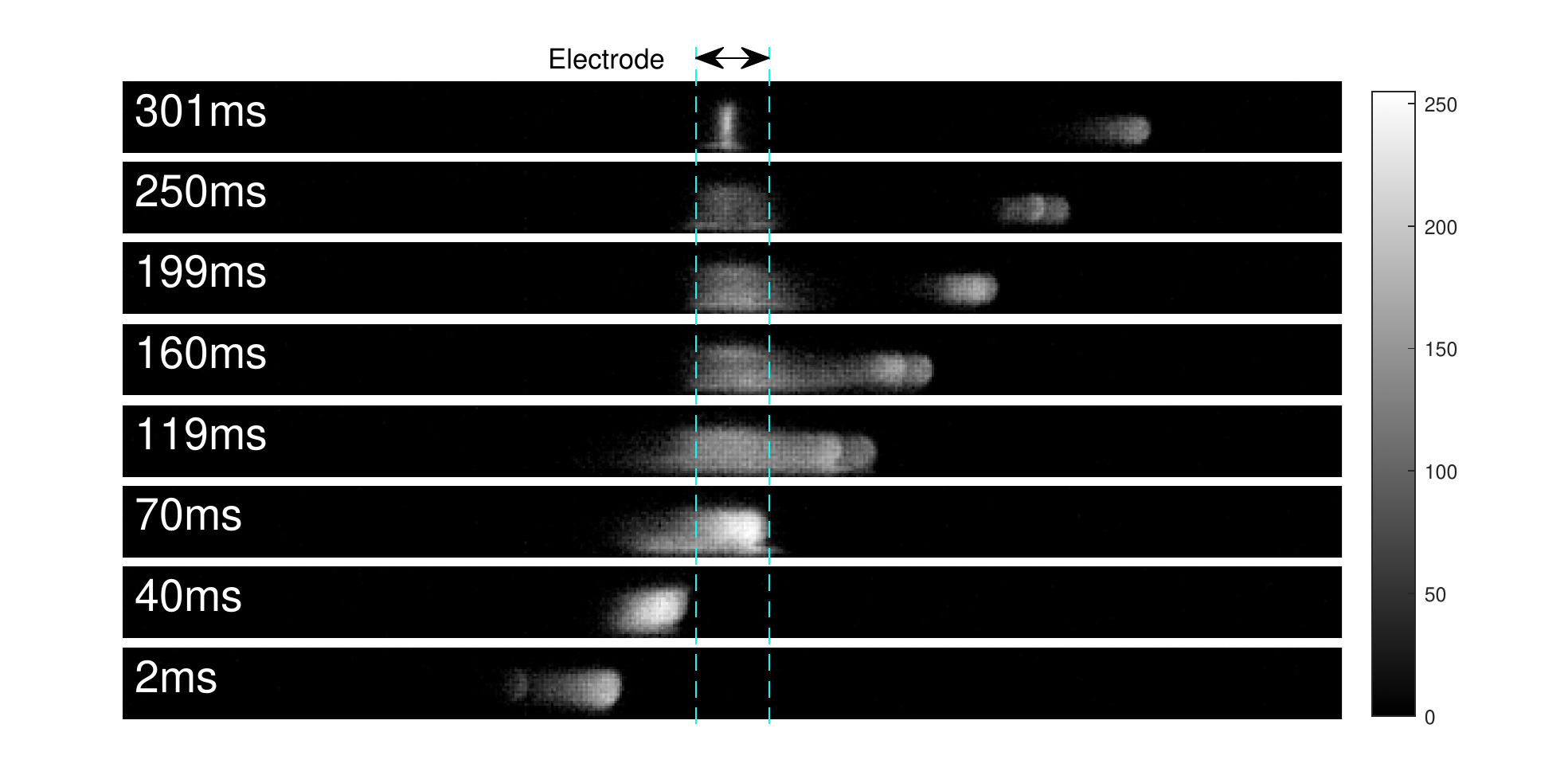}
    \caption{Evolution of the discharge when using the short electrode, video taken at 2kfps. Flame moves from left to right and channel height decreases from left to right. Time 0ms corresponds to the application of the first voltage pulse \adda{and pulsed voltage is applied during all frames. Conditions: f=10kHz,V=14.3kV, $\mathrm{\phi}$=1, Q=1000sccm, $N_p$=3300. Legend in colorbar is in arbitrary units.}}
    \label{fig:short_elec_evolution}
\end{figure}

An example set of images taken during a test with the short electrode is shown in figure~\ref{fig:short_elec_evolution}. \dela{The example  is  given  for  the  case  of  pulses  at  f=10kHz,V=14.3kV and flow conditions of $\mathrm{\phi}$=1, Q=1000sccm.}

The evolution with the short electrode shows some similarities with the long electrode. The initial behaviour is the same, with a period of increasing discharge size (70ms, 119ms frames). Unlike with the long electrode, there is a point (160ms in figure~\ref{fig:short_elec_evolution}) where the discharge and flame become separated. The discharge persists in the burnt gas directly underneath the electrode, but there is a region of hot gas that is not acted on by the high voltage between the discharge and the flame that emits no (visible) light. The long term behaviour is also different. With the long electrode, there was a transition to a filamentary regime immediately after the flame was quenched. With the short electrode, the discharge remains mostly uniform as it decreases in size and amount of energy deposited. In certain conditions, typically with higher frequency (8-10kHz) such as shown in figure~\ref{fig:short_elec_evolution}, there was an observed transition to a filamentary structure (see frame at 301ms). In other cases, typically with lower frequency, the discharge would extinguish directly from a contracted uniform region without first passing through the filamentary regime.

\subsection{Normalization by Flame Position}

\begin{figure}
    \centering
    \includegraphics[width=0.48\textwidth, trim={0.5cm 0 1cm 0},clip]{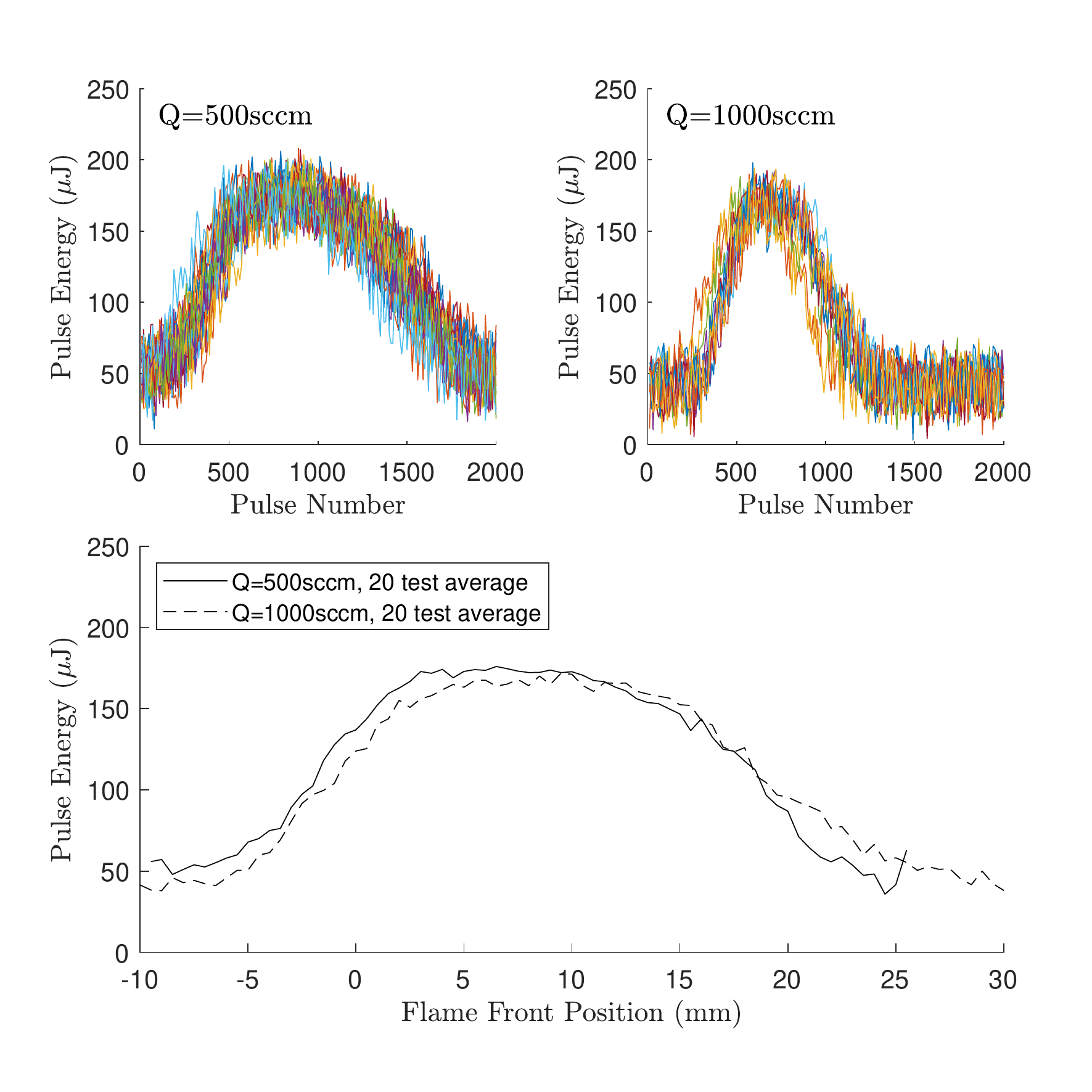}
    \caption{Normalization of the energy measurements by the flame position. \adda{f=8kHz, V=13.5kV, $\phi=1$ and N$_p$=2000 in both cases.}}
    \label{fig:flame_pos_norm}
\end{figure}

The flame speed in the lab frame is controlled by two parameters: the equivalence ratio and the flow rate of gas. Changing the equivalence ratio effects the quenching behaviour and burnt gas temperature; for the results presented in this work it was fixed at $\phi$=1.0. Changing $\phi$ affected the location of quenching, but not the overall behaviour; details can be found in the authors' previous work \cite{Pavan2021a}. The flow rate was observed to have a minor effect on the quenching distance; this is because the speed of the flame in the lab frame will influence how it couples to the walls of the chamber and the amount of stretch experienced in the boundary layers \cite{Ju2006}. Since this effect was minor, and the discharge is being applied before the channel has narrowed to the quenching distance of the flame, the general procedure used was to choose an equivalence ratio and then vary the flow rate to achieve the desired flame speed. Since the electrodes are fixed in the lab frame, changing the lab frame flame speed would modify the rate of discharge evolution. This is shown in figure~\ref{fig:flame_pos_norm}; both cases shown in the top two graphs were run with the same equivalence ratio, frequency and voltage but different flow rates. When the cases are plotted on the same graph, with the pulse number axis replaced by the flame position, the two cases perfectly overlap. This indicates that the evolution of the discharge is controlled entirely by the position of the electrode relative to flame front. This way of looking at the data also helps correct for small variations between tests that caused the flame to reach the electrode at slightly different times. The zero-reference for the flame front position axis is the edge of the electrode first encountered by the flame (the left side of the electrode in figure~\ref{fig:short_elec_evolution}).

Plotting against flame position also compensates for the changing flame speed (in lab frame) as it propagates through the channel. The flame slows as it passes through the tapered channel, both due to higher heat losses and because of a higher velocity of the unburnt gas. This causes the energy curves plotted on the pulse number axis in figure~\ref{fig:flame_pos_norm} to appear somewhat asymmetric, with the per-pulse energy rise rate faster than the decay rate. When plotted against the flame position, this asymmetry becomes much less pronounced because the effect of changing lab-frame flame speed has been removed.

\subsection{Energy Dependence on Frequency and Voltage}\label{sec:V_f_dependence}

\begin{figure}[]
    \centering
    \begin{subfigure}[]{0.45\textwidth}
        \centering
        \includegraphics[width=\textwidth]{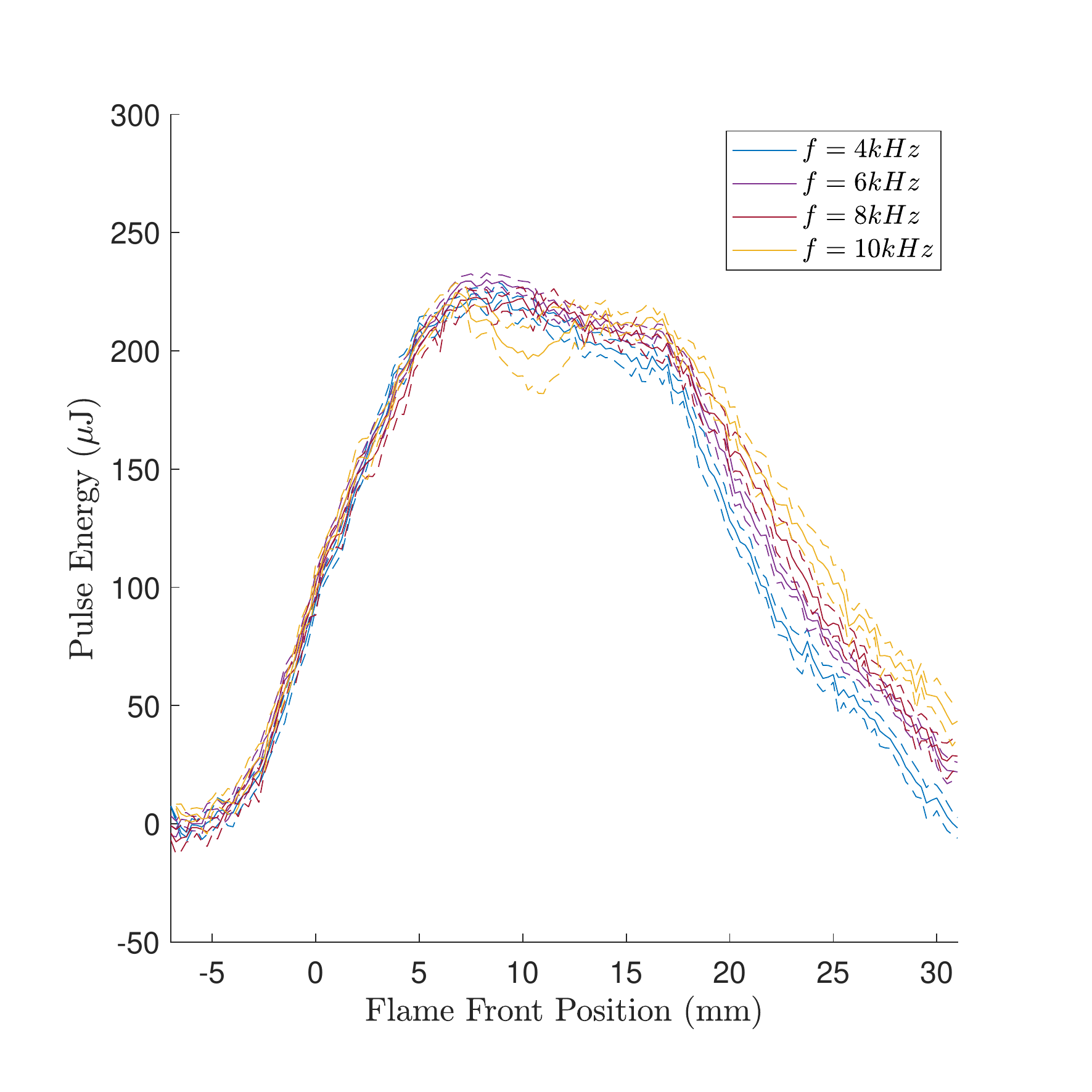}
        \caption{Frequency variation at fixed voltage V=14.3kV.}
        \label{fig:short_elec_nrg_freq}
    \end{subfigure}
    \begin{subfigure}[]{0.45\textwidth}
        \centering
        \includegraphics[width=\textwidth]{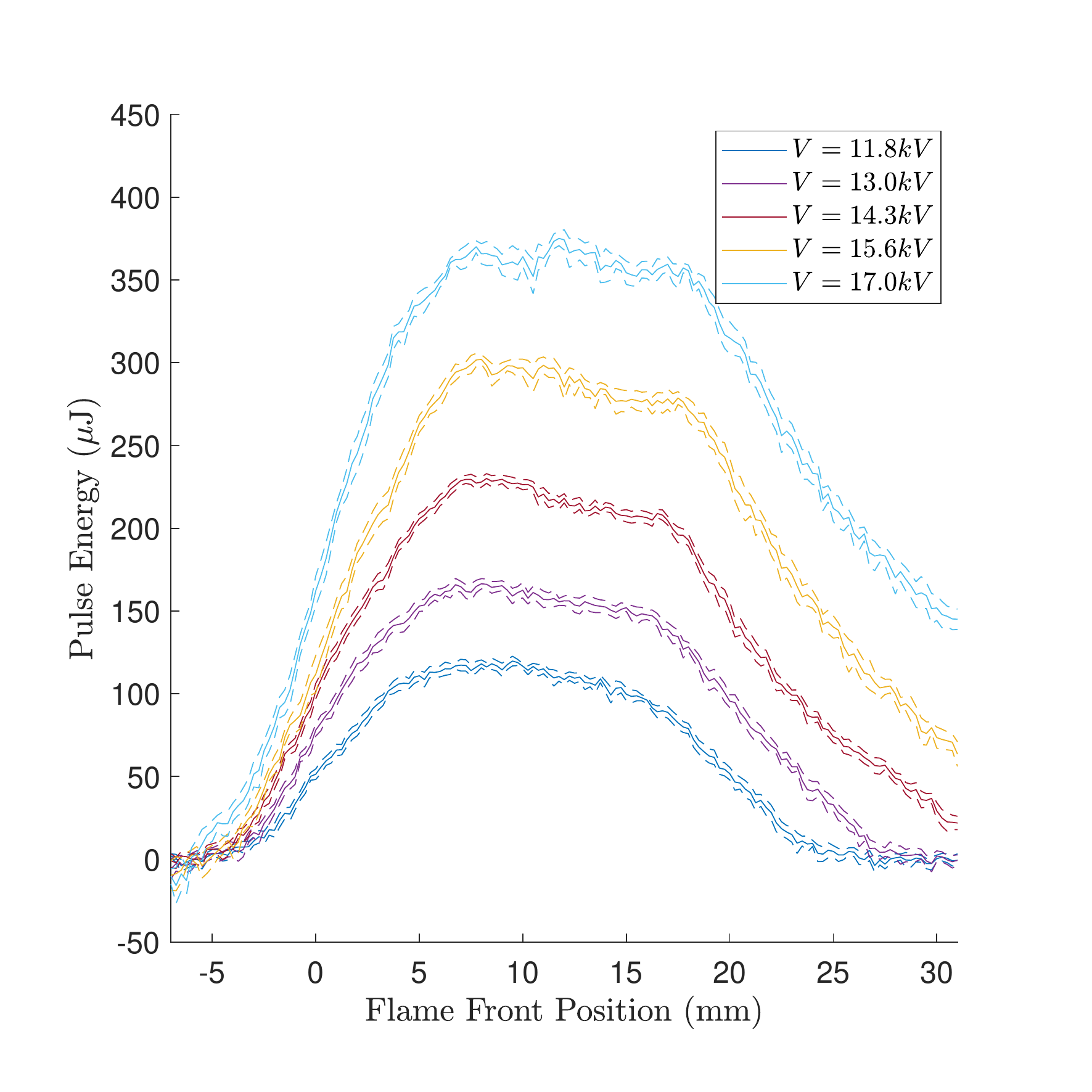}
        \caption{Voltage variation at fixed frequency f=6kHz.}
        \label{fig:short_elec_nrg_volt}
    \end{subfigure}
    \caption{Effect of varying the voltage and the frequency on the energy using the short electrode.}
    \label{fig:short_elec_nrg}
\end{figure}

The energy measurements were repeated for the short electrode. The equivalence ratio and flow rate were fixed and the frequency and voltage were varied. The results are shown in figures~\ref{fig:short_elec_nrg_freq} and \ref{fig:short_elec_nrg_volt}. The solid line represents the average of $\geq$20 tests and the dashed lines of the same colour are located at $\pm 2\sigma/\sqrt{N}$ where $\sigma$ is the standard deviation and $N$ is the number of tests. This is roughly a 95\% confidence bound on the mean energy. From figure~\ref{fig:flame_pos_norm}, the pulse to pulse energy varies but the mean energy across many tests is very consistent which explains the narrow bounds in figure~\ref{fig:short_elec_nrg}.

From figure~\ref{fig:short_elec_nrg_freq}, the frequency has almost no effect on the amount of energy deposited per-pulse. This implies that the gas condition at the start of each pulse is approximately the same regardless of pulse repetition frequency; quantities like the electron density reach a steady state in between pulses after relatively few pulses. When the flame is well past the electrode, there starts to be some deviation in the curves with the higher frequency depositing slightly more energy per-pulse than the lower frequency. The high speed videos indicated that this difference was caused by a slightly larger discharge region at the higher frequency. The effect of discharge region size is investigated more in the next section. The slight dip in energy when the flame is approximately 10mm past the start of the electrode for the 10kHz case corresponds to the flame being disrupted before continuing on, similar to the re-ignition phenomenon seen with the long electrode \add{ (see~\ref{sec:app_reinvig}).} \adda{A larger frequency having a greater disruptive effect was previously observed for this setup \cite{Pavan2021a}. The nanosecond pulser used in this work was limited to a maximum PRF of 10kHz, so higher frequencies could not be investigated, but the possible impacts are discussed further in section~\ref{sec:flow_effects}.}

The voltage on the other hand has a very significant effect on the energy deposited per-pulse, with the peak energy deposition doubling between 11.8kV and 14.3kV and almost doubling again when the voltage was increased to 17.0kV. The discharge also persisted for much longer after the flame had passed the electrode when the higher voltage was applied. At 11.8kV, even though pulses were still being applied, the discharge was no longer present after the flame front had moved 20mm beyond the end of the electrode, while at higher voltages a discharge was still visible and depositing energy well after the flame had passed and until the high voltage pulses were stopped. The peak per-pulse energy increases with the square of the applied voltage, as would be expected for energy stored in a capacitor.

\subsection{Energy Dependence on Discharge Region Length}

\begin{figure}[]
    \centering
    \begin{subfigure}[]{0.45\textwidth}
        \centering
        \includegraphics[width=\textwidth, trim={1cm 0 0 0},clip]{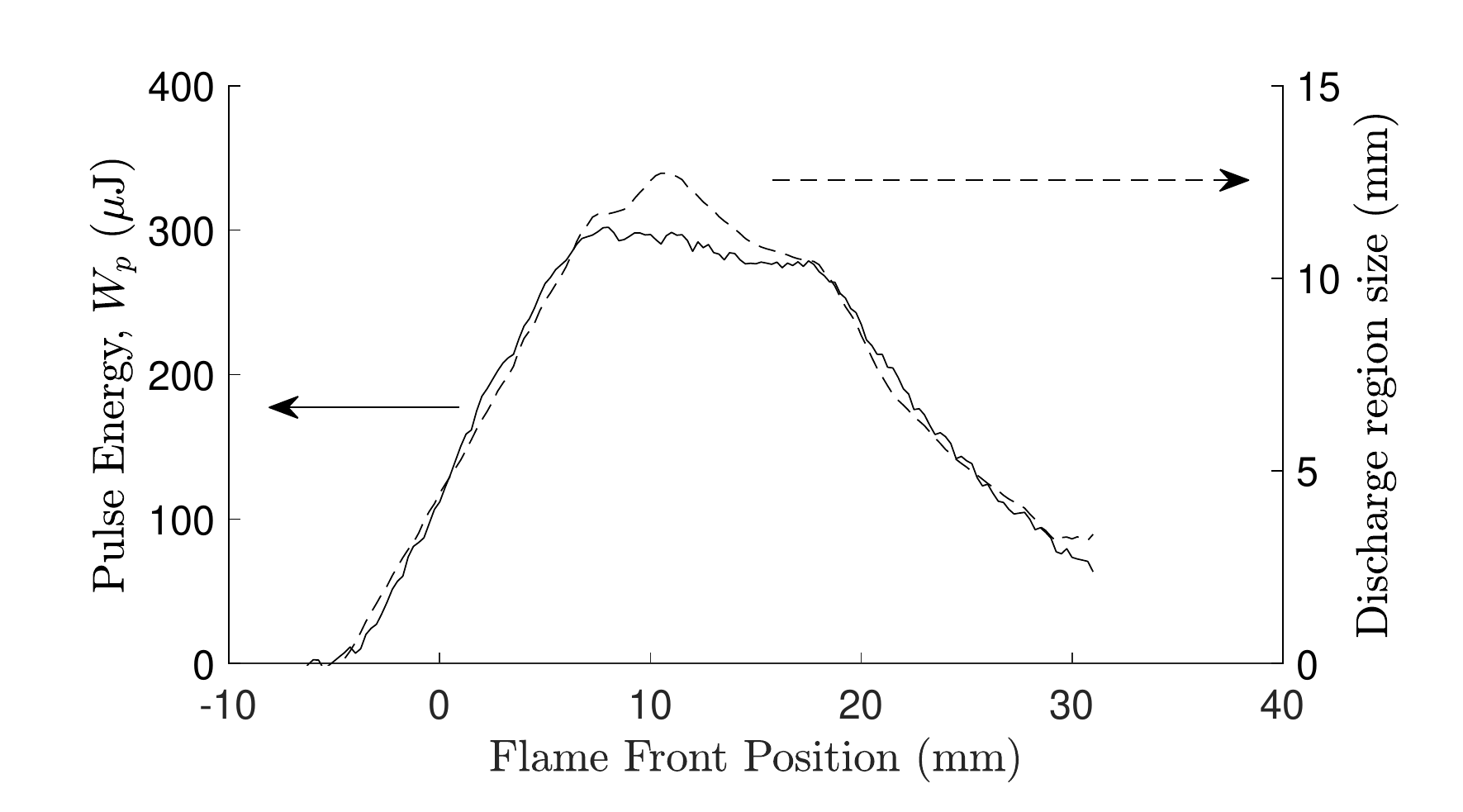}
        \caption{Size of discharge region and corresponding per-pulse energy.}
        \label{fig:nrg_vs_size}
    \end{subfigure}
    \begin{subfigure}[]{0.45\textwidth}
        \centering
        \includegraphics[width=\textwidth, trim={1cm 0 0 0},clip]{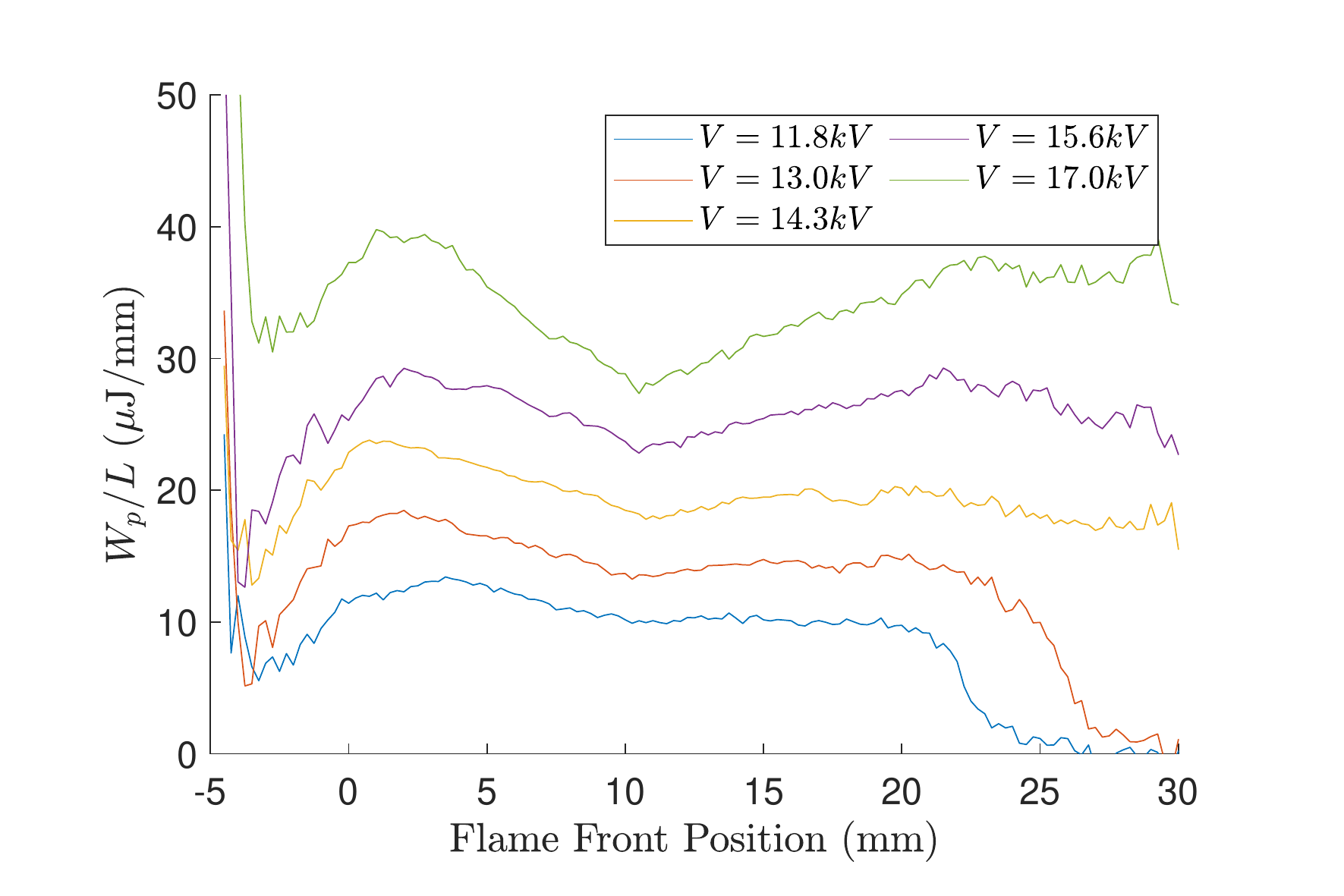}
        \caption{Per-pulse energy per unit length in flame propagation direction for the cases of figure~\ref{fig:short_elec_nrg_volt}.}
        \label{fig:nrg_per_length}
    \end{subfigure}
    \begin{subfigure}[]{0.45\textwidth}
        \centering
        \includegraphics[width=\textwidth, trim={1cm 0 0 0},clip]{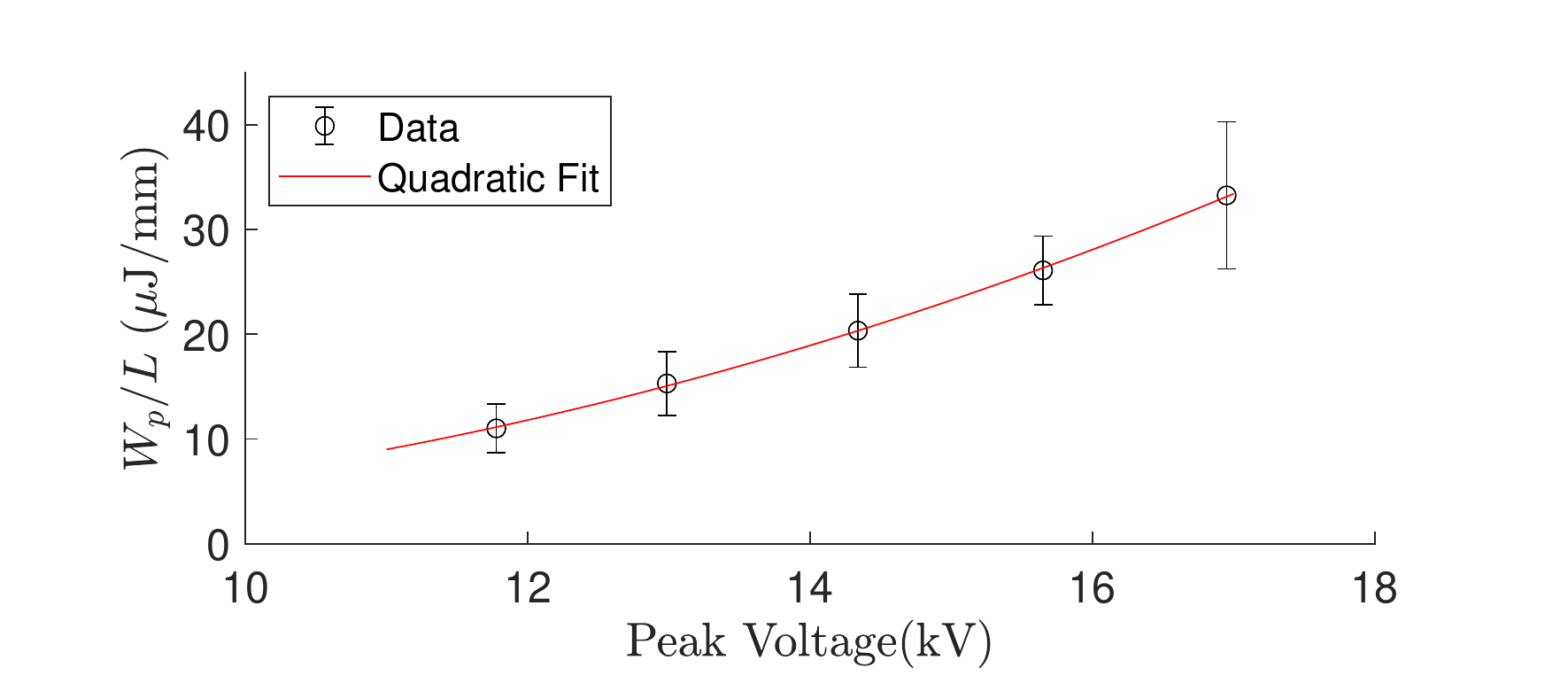}
        \caption{Quadratic fit of pulse energy per unit length for  figure~\ref{fig:nrg_per_length} between 0 and 20mm.}
        \label{fig:nrg_per_length_fit}
    \end{subfigure}
    \caption{Effect of discharge size on energy deposition.}
    \label{fig:discharge_size}
\end{figure}

The high-voltage electrode had a length in the flame propagation direction of 5mm, but the ground electrode extended over the full length of the test section. As a result, the discharge would often extend beyond the edges of the electrode due to electrode edge effects. This can be seen at 70ms and 119ms in figure~\ref{fig:short_elec_evolution}. After the flame passed, the discharge region would become localized underneath the electrode and then start to shrink in size as the gas cooled further. To account for the effect of discharge region size, figure~\ref{fig:short_elec_nrg} can be remade with the energy divided by the length of the discharge region. This will indicate whether the differences in energy deposition are due to changes in energy deposition density or changes in the size of the discharge region.
Figure~\ref{fig:nrg_vs_size} shows the evolution of discharge length along the electrodes and per-pulse energy as the flame passes under the electrode (average of 20 tests at f=6kHz, V=15.6kV). To account for the discharge extending beyond the extent of the electrode, the region used for discharge detection extended 5mm beyond the edge of the electrode in both upstream and downstream directions. This region was chosen because it was observed from the energy measurements that energy could begin to be deposited when the flame front was as early as 5mm upstream of the electrode. The presence of a discharge was determined visually from the high speed videos. The discharge and flame could be difficult to distinguish in the monochrome videos, but it was assumed that the discharge was always present on top of the flame when the flame was in the region below the electrode $\pm$ 5mm. This is a reasonable assumption, since after the flame had passed and the discharge and flame could be separated, the discharge persisted in this region. When the flame is present, the temperature and reduced electric field are higher, so the discharge is even more likely. The discharge also caused a distortion in the flame front shape that could further be used to confirm that it was overlapping with the luminous region of the flame. \add{This detection method is valid for uniform discharge mode only, which was the situation in all cases shown in figure~\ref{fig:nrg_per_length} except beyond 25mm.}
From figure~\ref{fig:nrg_vs_size}, it is observed that the per-pulse energy tracks with the discharge region size, suggesting that the energy density of the discharge is approximately constant in space. This is supported by figure~\ref{fig:nrg_per_length} which shows the per-pulse energy deposited per unit length of discharge \add{($W_p/L$)}. This value stays relatively constant for most of the discharge duration. As with the total per-pulse energy deposition, \del{the per-pulse energy deposition per unit length} \add{$W_p/L$} also increases with voltage. This indicates that the energy deposition with voltage increase seen in figure~\ref{fig:short_elec_nrg_volt} is indeed caused by the higher voltage, and not by a larger discharge region.

The relatively constant value of the per-pulse energy deposition per unit length at a given voltage indicates that the gas temperature has a minimal effect on the energy density deposited in any region of gas. This is significant, since the energy deposition pathways are highly dependent on the value of the reduced electric field \cite{Nagaraja2013}. This seems to indicate that by keeping the voltage constant, the same energy density would be deposited for different temperature gas, yet it would activate different kinetic mechanisms (determined by E/N). Nevertheless, the gas temperature does affect the size of the discharge region, since discharge can only be sustained in regions with elevated gas temperature. In particular, the discharge region size shrinks as the gas cools. \add{The value of $W_p/L$ as a function of peak voltage can be fit reasonably well using a second order polynomial as shown in figure~\ref{fig:nrg_per_length_fit}. The data points and error bars in this figure are the mean value and twice the standard deviation of the corresponding $W_p/L$ curves shown in figure~\ref{fig:nrg_per_length} sampled every 1mm between 0 and 20mm, the region where  $W_p/L$ is approximately constant.}
\subsection{Temperature Measurements using OES}\label{sec:OES_T_measure}
\del{\bf{4.5.1 Rotational-Translational Temperature}
.}
The RT temperature was measured simultaneously with the energy for all cases shown in figure~\ref{fig:short_elec_nrg} using the procedure described in section~\ref{sec:OES_methods}. In any individual test, the low number of exposures accumulated on the ICCD sensor resulted in a weak signal due to low amounts of light. To improve this, the spectra recorded for all 20 tests were combined, giving a much cleaner signal. To do this, the recorded spectra were placed into bins corresponding to a 1mm range of flame front locations. Recall that the size of the optical window and motion of the flame already cause some averaging of the spectra, so this binning procedure does not reduce spatial resolution. All the spectra in a bin were then summed, which greatly increased the signal strength. The temperature fitting was only performed on bins with at least 5 spectra, since those with less had too low a light level for a reliable temperature measurement.

\begin{figure}
    \centering
    \includegraphics[width=0.45\textwidth]{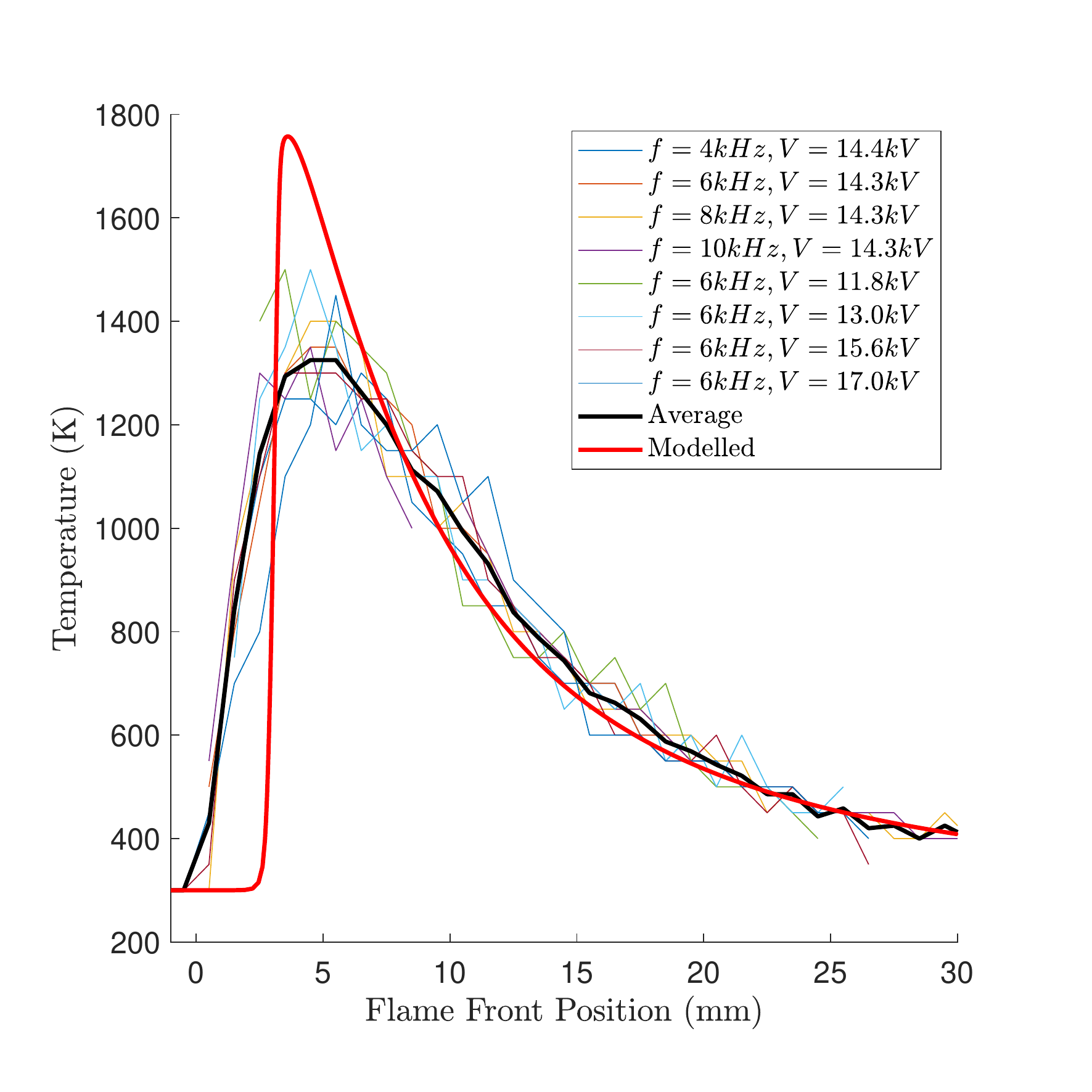}
    \caption{Measured rotational-translational temperature of the discharge for various cases. Modelled temperature of purely planar flame shown in red for reference.}
    \label{fig:T_profile}
\end{figure}

The measured temperature profile for the various tests is shown by the thin lines in figure~\ref{fig:T_profile}. All cases yielded a very similar temperature once the flame front had reached about 10mm past the beginning of the electrode. This suggests that any difference in heating caused by the discharge was small. This is compared to the maximum temperature rise that could be obtained if all the electrical energy was deposited as heat, calculated by equation~\ref{eqn:heating}.
\begin{equation}\label{eqn:heating}
    \Delta T_{\mathrm{max}}=\frac{W_p f}{\dot{m} c_p}
\end{equation}
Where the $\dot{m}$ is the mass flow rate and $c_p$ is the constant pressure heat capacity of the gas. Taking a value of $W_p=200\mu$J/pulse as typical energy deposition, f=6kHz, $\dot{m}=18.7\mu g/s$ and $c_p=1.1kJ/kgK$, the temperature rise caused by the discharge energy going entirely into heating is 58K, which is comparable to the RT reliable measurement precision. The undetectable temperature change between cases also allows for the average of the cases to be calculated, which is also shown in figure~\ref{fig:T_profile}.

The optical window was centered at 3.5mm from the left end of the electrode. When the flame front is near this region, there is a larger variation in the measured temperature. This can be attributed to two factors. First, the steep gradients in the temperature profile created by the flame meant that the discharge would not be occurring at a uniform gas temperature everywhere in the optical window and the coarse spatial resolution attainable in this setup could not capture the steep gradient. Second the flame is not a perfect planar front and non-uniformity across the cross-section, which varied between tests, also contributed to a wider variation. When the flame front had passed the optical window and the discharge was developing in the more uniform burnt gas region, the variation in the measurements was reduced. 

These factors that cause greater variation in the measured temperature when the flame front is near the optical window also cause the measured maximal temperature to be somewhat lower than what would be expected for a flame propagating in a rectangular section of comparable size. This is supported by modelling of the temperature profile in this configuration. A simple 1D model based on the free-flame solver in Cantera \cite{cantera} is used, with the finite size channel accounted for by adding a heat loss term corresponding to laminar flow in a rectangular cross-section pipe with constant wall temperature using a Nusselt number of 5.60 \cite{Bergman2011}. \add{Details of the model can be found in reference~\cite{Pavan2022}.} The modelled temperature and experiment agree well beyond 10mm, where the gas condition in the experiment is uniform and gradients are lower. Before this point, the experimentally measured temperature is considerably lower than what would be expected based on a purely planar flame front. It is therefore concluded that the temperature measured when the flame front is beyond 10mm from the electrode start are reasonably accurate, while the temperature measured when the flame front is before this point corresponds to a spatial average in the non-uniform environment of the observed section.

\del{
\bf{4.5.2 Vibrational Temperature}
.}
\section{\adda{Discussion: Discharge Dynamics and Regime Transitions}}
\subsection{\adda{Role of E/N}}\label{sec:E_N_effects}

Discharge mechanics are largely controlled by the reduced electric field, and therefore it is useful to relate the observed discharge evolution to this parameter. For the setup in this work, the voltage across the air gap ($V_g$) is related to the applied voltage ($V$) by equation~\ref{eqn:gap_voltage}\add{, which is the equation for three sequential parallel plate capacitors, dielectric-gas-dielectric, with the first and last identical.}
\begin{equation}\label{eqn:gap_voltage}
    \frac{V_g}{V}=\left[1+2\frac{\epsilon_g d_q}{\epsilon_q d_g}\right]^{-1}
\end{equation}
Where $\epsilon$ is the (relative) dielectric constant (1 for the gas, 3.8 for quartz), $d$ is the thickness of the layer and the subscripts $q$ and $g$ refer to the quartz and the gas respectively. For the location of the short electrode, this gives a ratio of the gas gap voltage to the applied voltage of approximately 0.75. Prior to any ionization, the electric field in the gap is given by $V_g/d_g$, however as the discharge evolves charge will build up on the surfaces of the dielectrics and sheaths will form in the discharge. Modelling of the reduced field as a function of position in a NRP DBD discharge has been performed by \cite{Adamovich2009, Nagaraja2013} \add{and recently developed experimental techniques have begun to allow it to be probed directly \cite{Orr2020,Rousso2020}.}

\begin{figure}
    \centering
    \begin{subfigure}[]{0.45\textwidth}
        \centering
        \includegraphics[width=\textwidth, height=6.5cm,trim={0.5cm 0.2cm 1cm 1cm}, clip]{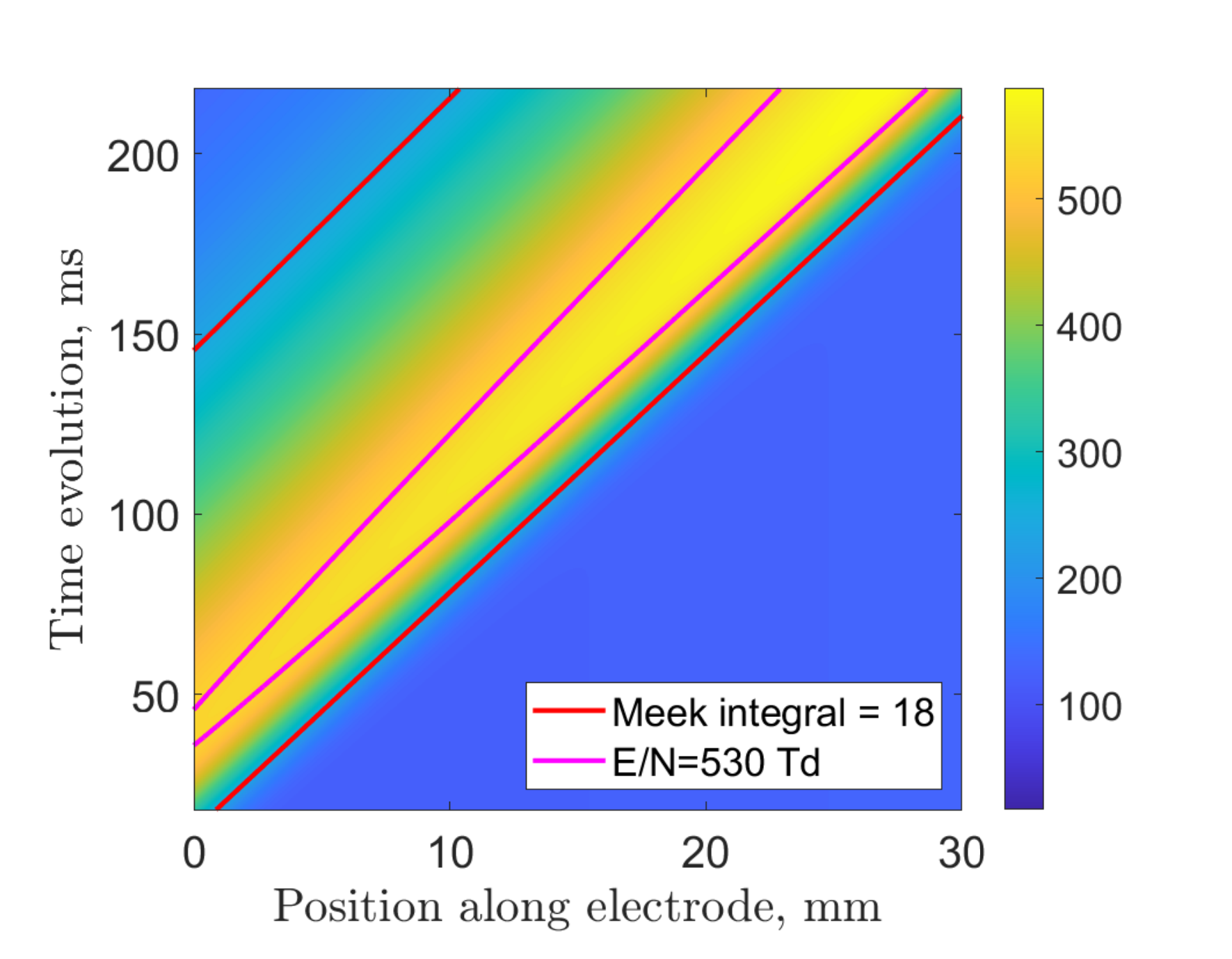}
        \caption{Reduced field (Td) as a function of space and time. Contours corresponding to Meek breakdown criteria and E/N=530Td highlighted. Model is for the long (30mm) electrode with a peak voltage of 13kV.}
        \label{fig:E_N}
    \end{subfigure}
    \begin{subfigure}[]{0.45\textwidth}
        \centering
        \includegraphics[width=\textwidth, height=6.5cm, trim={1cm 1cm 1.5cm 1cm}, clip]{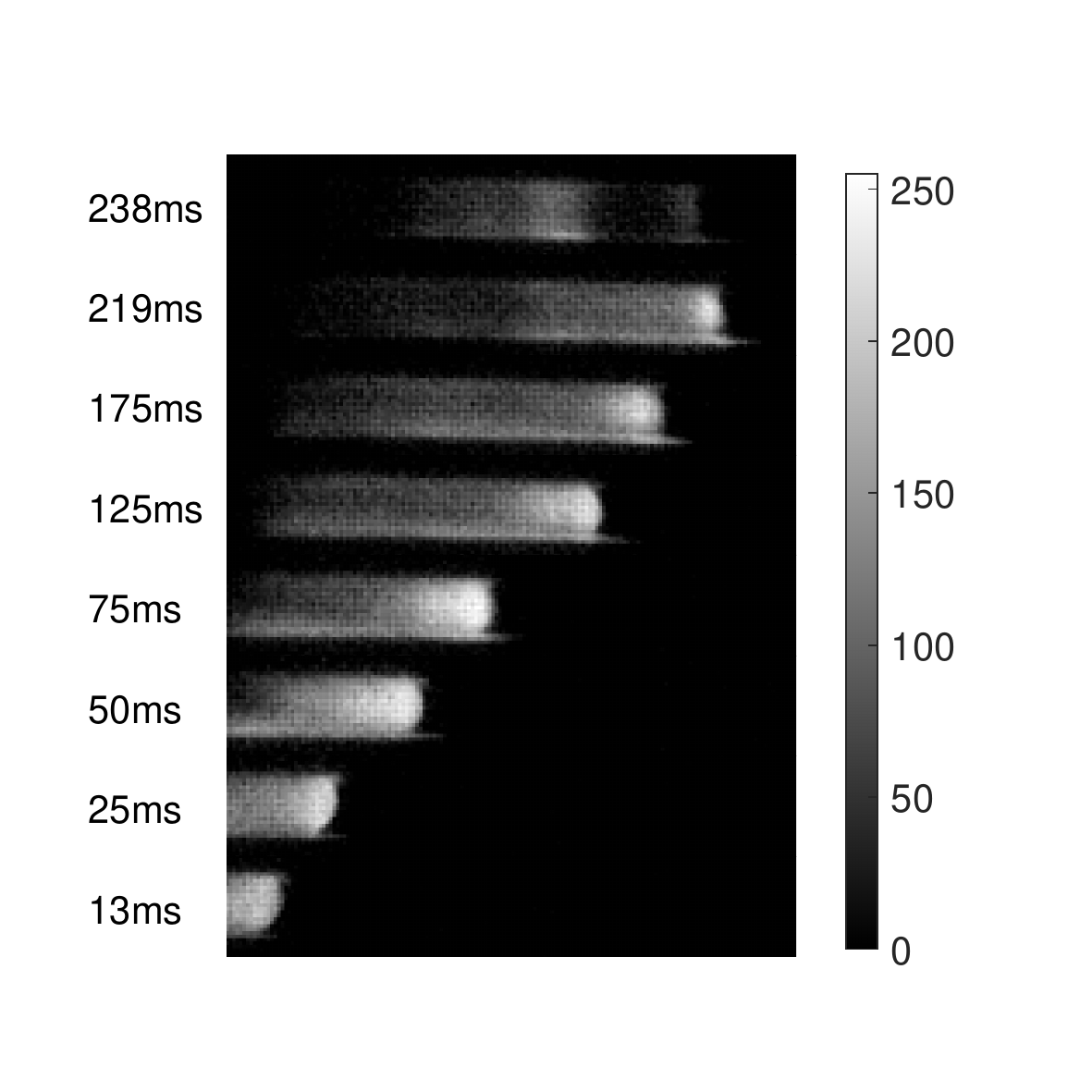}
        \caption{Image of flame evolution under long electrode with f=8khz, V=13kV, $\phi$=1.0, Q=1000sccm, Np=2000. Legend for colorbar is in arbitrary units.}
        \label{fig:flame_appearance_E_N}
    \end{subfigure}
    \caption{\adda{Comparison of reduced electric field to flame appearance.}}
    \label{fig:E_N_analysis}
\end{figure}

Useful information can be obtained just based on the reduced electric field calculated from the applied voltage (and not including the effect of space charge, but including the presence of the dielectrics); for example the work by Liu et al. on regime transitions identifies a threshold of 130kV/cm at atmospheric pressure measured based on the applied voltage, corresponding to 530Td \cite{Liu2018,Liu2019,Liu2014}. It should be noted that these works look at a single nanosecond pulse, and not a NRP system, which will have additional effects caused by the accumulated pulses. \del{The reduced field based on applied voltage for both the experimental and modelled temperature profile in this work at two different applied voltages is shown in figure~\ref{fig:E_N}}.\add{The reduced field based on applied voltage and the average measured temperature from figure~\ref{fig:T_profile} is shown in figure~\ref{fig:E_N} as a function of position and time. In this figure, position=0mm corresponds to the left edge of the long electrode (gap height 3.4mm) and position=30mm corresponds to the right edge (gap height=2.9mm) and a flame speed of 15cm/s (lab frame) has been assumed. Contours corresponding to two critical thresholds are highlighted\dela{: the}\adda{. The first is a} breakdown threshold based on the Meek integral condition for streamer formation\dela{(a metric based on the accumulated electrons produced in an electron avalanche}\adda{, equation~\ref{eqn:Meek}, which is a metric characterizing the increase in the number of electrons in an electron avalanche \cite{Fridman2011, Raizer1991}.} \dela{, and}
\adda{\begin{equation}\label{eqn:Meek}
    K=\int_0^\ell \alpha\left(E/N\right)dx
\end{equation}
x is the coordinate along the electron avalanche direction, $\ell$ is the length of the avalanche and $\alpha$ is the Townsend ionization coefficient which is a function of the reduced electric field. A value of K=18-20 typically corresponds to the formation of a streamer \cite{Fridman2011, Raizer1991}.}  

\adda{The other is metric is }the threshold for uniform discharge proposed by Liu et al. for a single pulse.}. In figure~\ref{fig:flame_appearance_E_N}, several images of the flame are shown in a test with the same conditions as figure~\ref{fig:E_N} (V=13kV, 30mm electrode, flame speed $\approx$15cm/s in lab frame). By comparing figure~\ref{fig:flame_appearance_E_N} to figure~\ref{fig:E_N}, it is clear that the discharge is present in a region approximately corresponding to where the Meek integral is $\geq$18.

\del{The reduced field near the reaction zone of the flame front exceeds the uniform threshold in all cases shown in figure~\ref{fig:E_N}, but based on the 530Td threshold there should be a regime transition when the leading edge of the flame is around 10mm past the start of the electrode. This is not what was observed; the transition happened when the flame was much further along, if it even happened at all. In the next section, an alternate mechanism for regime transition is discussed, one that takes into account the repetitive nature of the NRP discharge.}\add{The Meek integral criterion, usually used as a threshold for streamer formation but applicable here as a metric of the electron avalanche producing a sufficient quantity of electrons to sustain the discharge, does a pretty good job of estimating when the discharge will occur. Note the region at the top left of figure~\ref{fig:E_N} where the criterion is not met and compare this to \dela{figure~\ref{fig:long_elec_evolution} at t=175ms}\adda{the dark region at the top left of figure~\ref{fig:flame_appearance_E_N}} when the discharge has disconnected from the left edge of the electrode. However, the criterion for uniform to filamentary transition proposed by Liu et al. does not seem to be applicable here, as the discharge appears uniform for a wide range of E/N. The most likely explanation is that the accumulated affects of many pulses change the dynamics of the transition, which is discussed further in the following section.}

\subsection{\dela{Mechanisms Contributing to Regime Transitions}\adda{Role of Gas Properties and the Thermal-Chemical Instability}}\label{sec:gas_effects}
\dela{
Different behaviour was seen for the short and long electrodes, that can not be easily attributed to variations in E/N. For the long electrode, the flame would typically quench before exiting the inter-electrode region. This quenching causes a rapid drop in gas temperature, and the discharge region quickly contracts and forms filamentary structures before it completely extinguishes. With the short electrode, the flame would pass through the electrode the majority of the time and the gas under the electrode cooled much more gradually. In this case, the transition to filaments was only observed at the highest frequencies and voltages, when the discharge persisted long past when the flame had passed. At lower frequencies, the discharge would slowly contract and ultimately extinguish directly from the uniform regime, without any clear microdischarges having been formed. Even in cases where the microdischarges do form with the short electrode, they happen at a much lower reduced field than would be expected purely based on the 530Td calculated from applied voltage criteria. Looking at figure~\ref{fig:short_elec_evolution}, the discharge is still uniform at 200ms when the flame is 20mm past the start of the electrode. \del{From figure~\ref{fig:E_N},} \add{Based on the measured temperature, however,} the reduced field is below 300Td, much less than the transition threshold proposed in references \cite{Liu2018,Liu2019,Liu2014}.}

Another factor affecting the discharge uniformity are the ionization properties of the gas \cite{Adamovich2009, Gadkari2017, Rousso2020}. \add{Certain combustion products, such as NO  \cite{Adamovich2009}, have a lower ionization energy than O$_2$ and N$_2$ and have been proposed to promote instability. At the same time, the presence of water vapour in the burnt gas may have a stabilizing effect due to its high attachment rate \cite{Zhong2019}, which could explain the long-lived uniform discharge observed in this work. The existing population of electrons (as a result of previous pulses) will also have an effect, with higher residual electrons being linked to earlier breakdown \cite{Huang2014} and an electron energy distribution function closer to Maxwellian \cite{Huang2018}.} \adda{Another possibility, discussed in \cite{Zhong2019, Rousso2020} is heat release or absorption by chemical reactions in the flow triggering the transition between discharge regimes. Filaments are often seen after the flame has quenched. It is possible that partial oxidation of fuel that has diffused into the hot gasses is occurring and that this is contributing to the regime transition. The diagnostics used in this work do not allow to quantify the impact of chemistry on the regime transitions, but a number of recent papers have focused on this contribution \cite{Zhong2019, Rousso2020}.}

\subsection{\adda{Role of Fluid Residence Time and Pulsation Frequency}}\label{sec:flow_effects}

\add{For a flowing gas, it may not be the pulse repetition rate on its own but rather how this timescale compares to the gas residence time that matters \cite{Khomich2016, Zhang2021}.}\del{The effect of gas residence time with the NRP DBD offers a better explanation in this case.} Zhang et al. \cite{Zhang2021}, observed that a uniform discharge could be sustained with a reduced field of 150Td (neglecting space charge) and a flow residence time to pulse repetition period ratio ($\tau_{\mathrm{res}}/\tau_{\mathrm{PR}}$) of\del{$\mathcal{O}(10)$}\add{order 10}, while a filamentary structure was more likely with ratios of \del{$\mathcal{O}(\geq100)$}\add{100 or greater} at the same electrical conditions. Khomich et al. \cite{Khomich2016} proposed a condition based on the fluid motion of the gas between pulses transporting the gas far enough that residual inhomogeneities from the previous pulse would not be amplified, estimating a critical distance of 0.9mm. 

\begin{figure}
    \centering
    \includegraphics[width=0.45\textwidth, trim={1cm 1cm 1cm 0cm},clip]{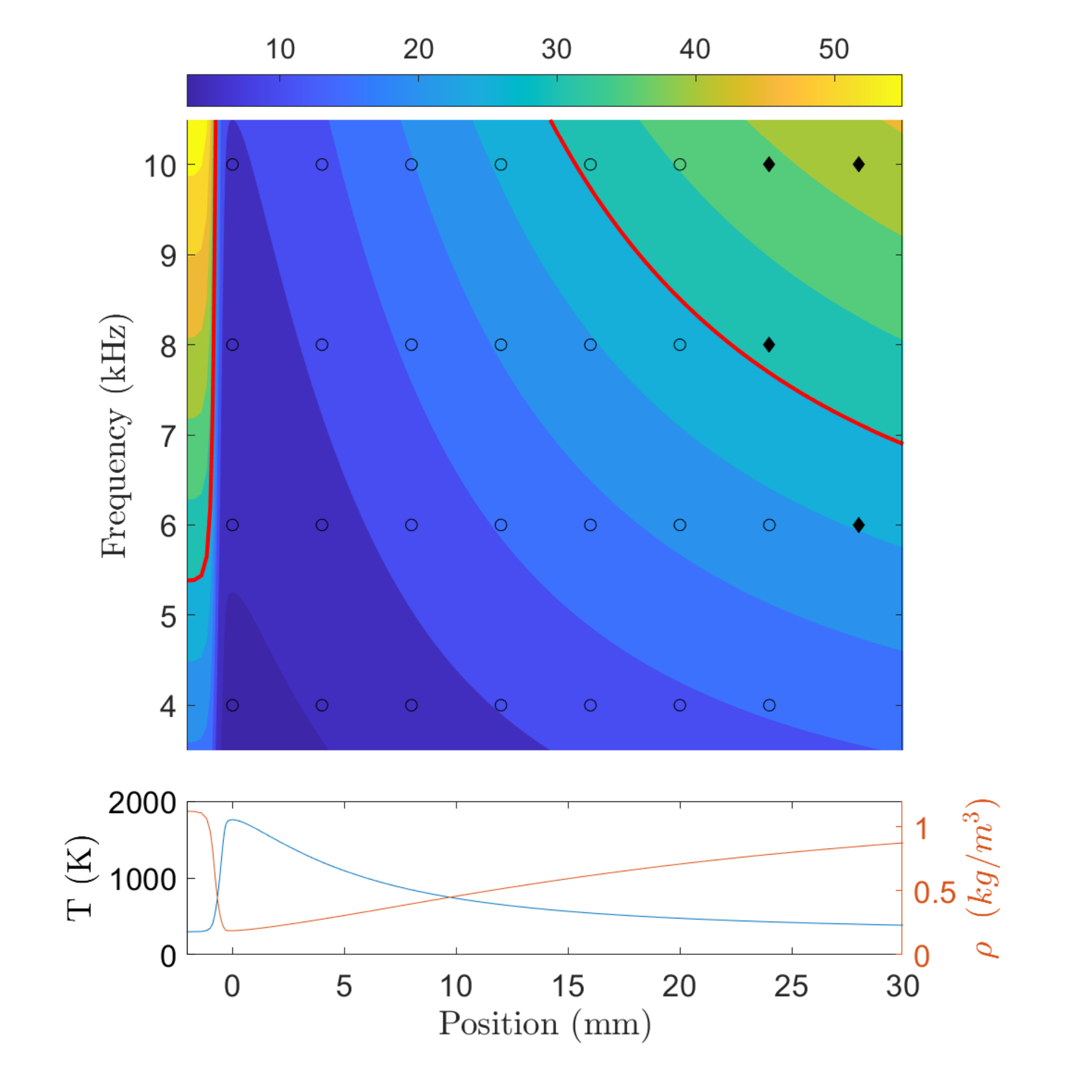}
    \caption{Contours of \dela{$\tau_{\mathrm{res}}/\tau_{\mathrm{PR}}$}\adda{$\tau$} (non-dimensional) for short electrode with modelled flame temperature and density shown for reference. The thick red line indicates a ratio of \dela{150}\adda{30}. Experimental data is shown as $\circ$ (uniform regime) and $\blacklozenge$ (filamentary regime).}
    \label{fig:regime_transition}
\end{figure}

\adda{The work of Zhang et al. used the electrode size as the characteristic dimension for calculating residence time, while the work of Khomich et al. used the discharge filament size. The latter metric will be used in the following analysis, since it will be electrode area-independent. The geometry of the discharge cell means that the mass flow rate will be constant. There is a slight increase in the height of the channel h, and the channel has a constant span, w. Thus, the flow rate can be calculated by equation~\ref{eqn:flow_velocity}
\begin{equation}\label{eqn:flow_velocity}
    u(x)=\frac{\dot{m}}{w} \frac{1}{\rho(x) h(x)}=\frac{\dot{m}R}{w P }\frac{T(x)}{h(x)}
\end{equation}
where u is the average flow velocity, x is the coordinate along the flame propagation direction, $\dot{m}$ is the mass flow rate and $\rho$ is the density, R is the gas constant, P is the pressure and T is the temperature. Since the gas composition is dominated by nitrogen, R changes very little between the burnt and unburnt mixture. Equation~\ref{eqn:timescale_ratio} is used to get a non-dimensional ratio, $\tau$ of flow timescale to discharge timescale
\begin{equation}\label{eqn:timescale_ratio}
    \tau=\frac{f_{p} d}{u(x)}=\frac{w P f_{p} d}{\dot{m}R}\frac{h(x)}{T(x)}
\end{equation}
where $f_{p}$ is the pulse repetition frequency and d is the characteristic length of the instability leading to filament formation.
}
\dela{The condition of Zhang et al. agreed well with the results of this experiment as shown in figure~\ref{fig:regime_transition}. The contour plot shows the value of $\tau_{\mathrm{res}}/\tau_{\mathrm{PR}}$. The threshold for transition is taken to be 150 and is marked in red.}\adda{Taking a fixed observation location where the channel height is 3.1mm and assuming a characteristic filament length scale of d=1mm, the contours of figure~\ref{fig:regime_transition} are constructed.} In the experiment, transition to filamentary regime in the short electrode was only observed at high frequency once the flame was well past the electrode (see figure~\ref{fig:short_elec_evolution}), which corresponds to the top right region of the contour plot. At lower frequencies and voltages, the discharge would extinguish before a regime transition was observed. Observations from the experiment are shown on the contour plot as  $\circ$ for uniform regime and $\blacklozenge$ for filamentary regime. The transition observed in the experiment is reasonably well explained by the residence time to pulse repetition period ratio \adda{taking 30 as an approximate value for the transition}.
The transition to filamentary discharge with the long electrode can also be explained in this way, since with the long electrode the flame would generally quench before exiting the electrode. In that case, the flow rate would drop down to the oncoming gas flow rate (the far left side of the contour plot, ahead of the flame front) regardless of temperature because the force generated by the expanding gas no long exists. This may explain why the long electrode almost always displayed transition to filamentary regime when the flame extinguished. In cases where the flame nearly quenched and then reignited the discharge \add{(see~\ref{sec:app_reinvig})}, the discharge always went back to a uniform mode. \dela{Khomich et al.'s condition proved too conservative to explain the observations of this experiment, with the gas only needing to move \del{$\mathcal{O}(10\mu m)$}\add{on the order of $10\mu$m} to suppress filamentary transition. The difference with Khomich's work is that they use a traditional DBD with square pulses rather than the NRP system used in this work, and so comparison to their threshold is less meaningful. Regardless, the observations of this work support the theory that regime transition can be suppressed at flow rates that are high enough to disrupt inhomogeneity amplification on the dielectric surfaces.} \adda{As noted in section~\ref{sec:V_f_dependence}, the maximum frequency that could be used was 10kHz due to equipment limitations. It is expected that a higher PRF would cause earlier transition to a filamentary regime, but further experiments are necessary to confirm this.}

\section{Conclusions}

This work investigated the evolution of a NRP DBD discharge as it interacted with a transient laminar flame on the time scales of the flame propagation. Two configurations were tested; a long electrode was used to look at the evolution of the discharge over a large region of space as the flame passed under it, and a small electrode was used to look at the discharge evolution in a local region as the gas condition changed. The discharge was strongly affected by the presence of the flame, and the main observations can be summarized as:
\begin{itemize}
    \item Discharge only occurred in the hot gases produced by the passage of the flame (including possibly the reaction zone).
    \item The per-pulse energy deposited by the discharge was directly related to the size of the discharge region and the applied voltage, and was unaffected by the frequency or flow rate of gas.
    \item Transition from uniform to filamentary discharge mode was only observed in the combustion products after the flame had quenched or travelled a significant distance away from the discharge region. The relative time scales of gas motion and pulse repetition rate can be used to explain this transition. The reduced electric field alone can not explain the observations.
\end{itemize}
The coupled evolution of an NRP DBD discharge with a transient flame has important implications for PAC systems. This work demonstrates that systems involving transient combustion processes must consider the changing gas environment and the effect it will have on the discharge behaviour, energy deposition and energy pathways.

\section*{Acknowledgments}
This work has been supported by Lockheed Martin Corporation (LMC) as part of its University Research portfolio, within the program managed by Dr. John Rhoads and Mr. Jeffrey A. Mockelman acting as technical monitor. The authors also acknowledge support from the Office of Naval Research (ONR) through the Young Investigator Program, under Award Number N00014-21-1-2571, within the program managed by Dr. Steven Martens.

\appendix
\add{
\section{Energy Calculation}\label{sec:app_energy}
The energy calculation is performed by integrating the product of current and voltage by equation~\ref{eqn:nrg_calc}. The mismatched impedance in the system caused many pulse reflections of decreasing amplitude after the initial pulse. These reflections continued for approximately 800ns before a switch in the pulser damped them rapidly. Most energy deposition to the discharge occurred during the initial pulse, but to get a proper energy measurement the signals had to be integrated until about 700ns to account for the charging and discharging of the capacitor. A typical trace of voltage, current and the integrated product (energy) for a case with and without visible discharge is shown in figure~\ref{fig:waveforms} with the current shown in detail in figure~\ref{fig:current_closeup}. Three cases are shown with no visible discharge, a low energy discharge and a high energy discharge. In these cases, the energy deposited by the pulse is recorded as $4\mu J$, $146\mu J$ and $326\mu J$ respectively. All three sets of waveforms were generated during the same test, with the different behaviour caused by the flame passage. The current waveforms show more reflections than the voltage waveform because it was located on the low-voltage side of the discharge gap, a necessity to avoid sparks to the Pearson coil. Downstream of the discharge gap, there are superimposed waves coming from the the current flowing through the discharge gap from the high voltage side (capacitive and conductive components) and current reflected from the system ground connection, which is why the frequency of the current oscillations is greater than the voltage oscillations (which travel between the discharge gap and the pulser), however these reflected pulses still represent a real current flowing towards the discharge gap.
\begin{figure}
    \centering
    \begin{subfigure}[]{0.45\textwidth}
        \includegraphics[width=\textwidth, trim={1cm 1.5cm 1cm 1cm},clip]{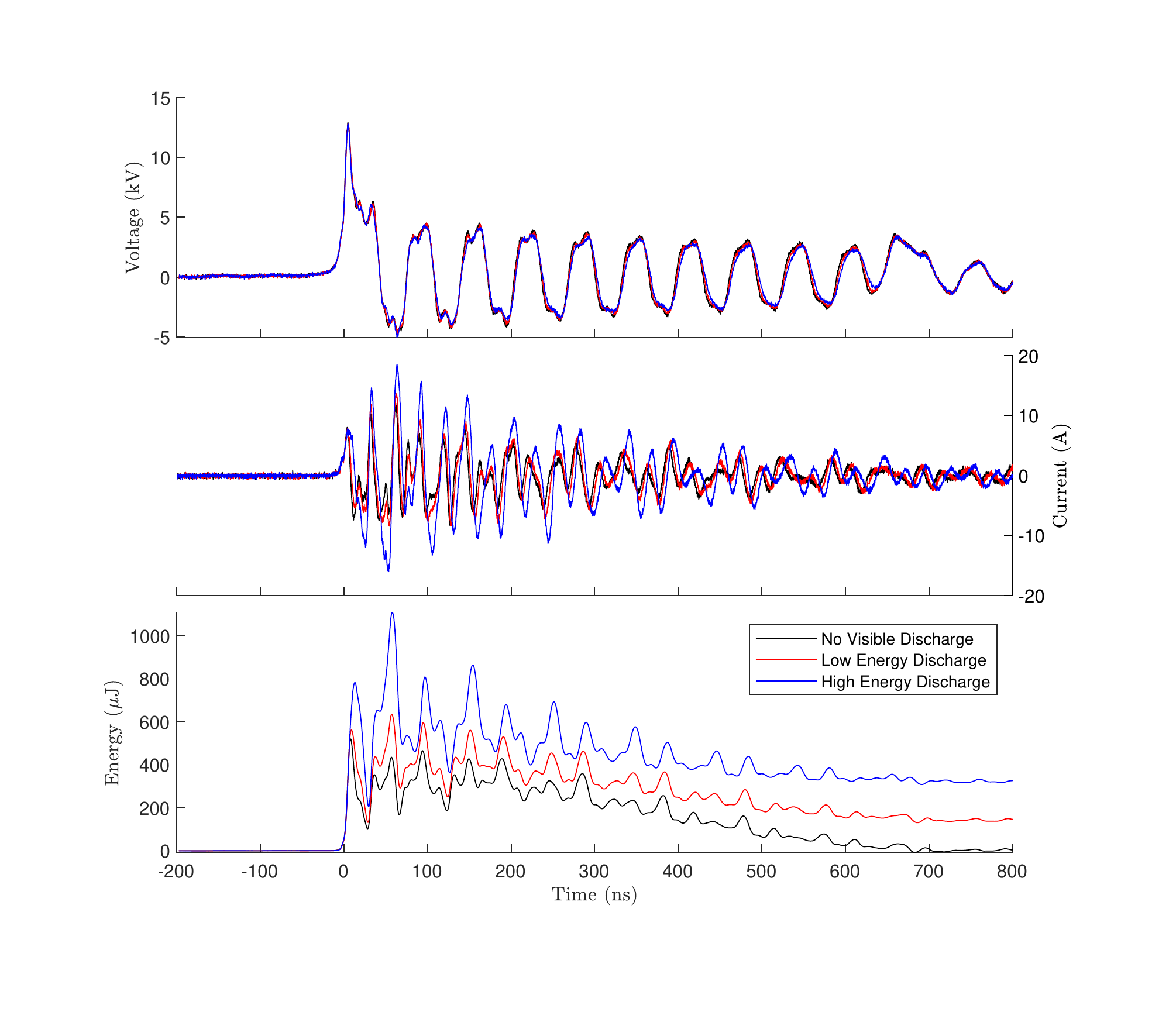}
        \caption{Voltage and Current waveforms and energy measurement for three discharge cases.}
        \label{fig:waveforms}
    \end{subfigure}
    \begin{subfigure}[]{0.45\textwidth}
        \centering
        \includegraphics[width=\textwidth,trim={1cm 1.5cm 1cm 1cm}, clip]{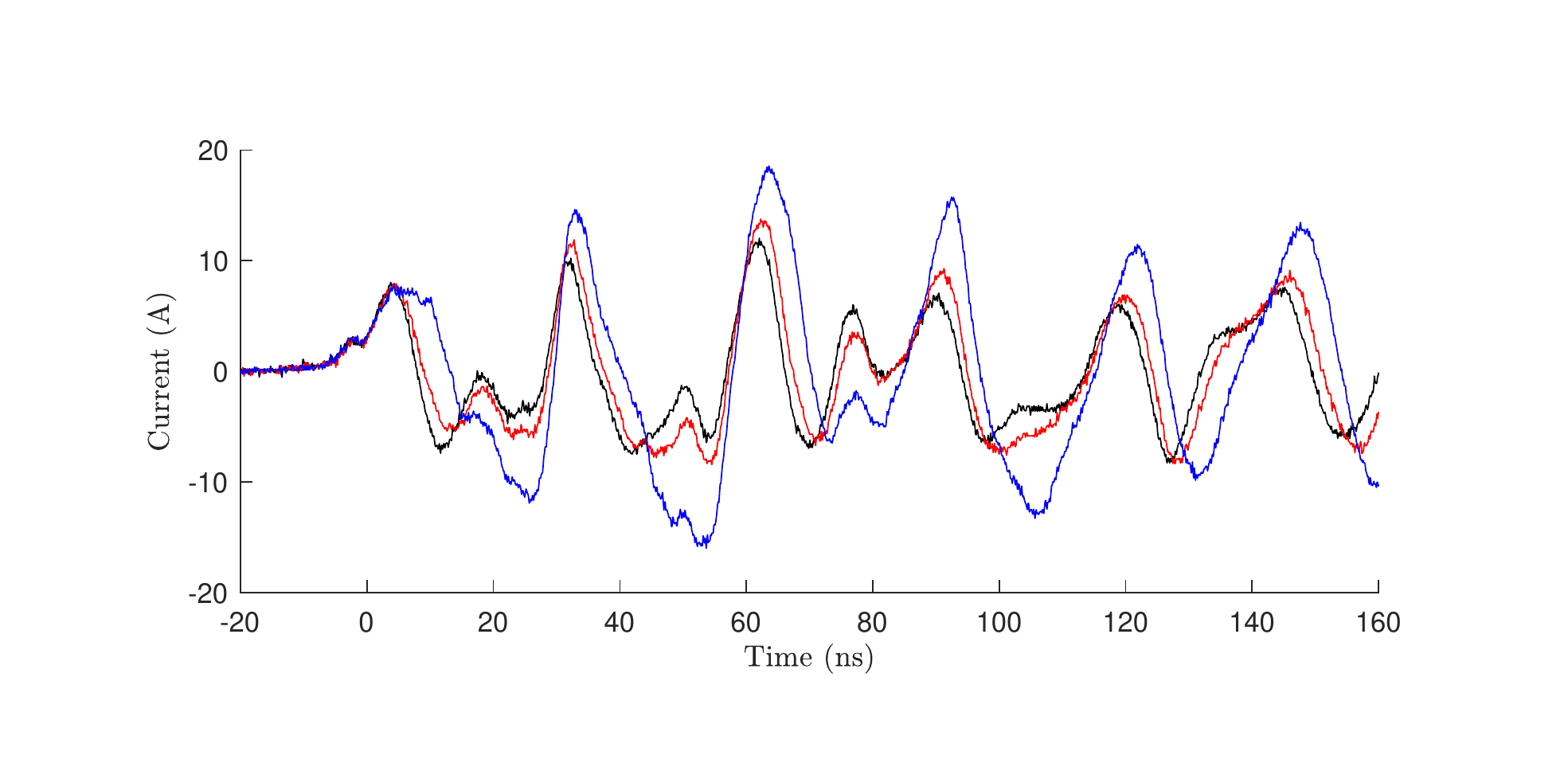}
        \caption{Zoomed in view of current waveforms during the initial pulse and first two reflected voltage pulses.}
        \label{fig:current_closeup}
    \end{subfigure}
    \caption{Example waveforms and energy measurements. \adda{Conditions: f=8khz, V=13.0kV, $\phi$=1.0, Q=1000sccm, N$_p$=3200. No discharge, low energy discharge and high energy discharge are pulse 50, 450 and 1200 respectively.}}
    \label{fig:Waveforms_parent}
\end{figure}
The nanosecond timescale of the voltage rise means that signal travel time delays are significant and must be accounted for to get accurate energy measurements. The voltage probe was located on the high voltage side of the test section, and the current probe on the current return side since undesirable coronas formed between the probe and the wire when it was placed on the high voltage side. This spatial separation, combined with different length cables from the probes to the oscilloscope, led to a lag between the current and voltage signals. To account for this lag, the high voltage was pulsed at a low enough voltage that no discharge would form; this resulted in the circuit behaving as a pure capacitor. In this case, no energy should be dissipated in the load, so the lag was set so that the total energy dissipated in the load is 0.  This lag was on the order of a few nanoseconds, and could vary based on small changes to the setup, but has a significant impact on the energy calculation. A demonstration of this effect is shown in figure~\ref{fig:lag_demo} for a case without breakdown in the discharge gap. The red curve shows the energy calculation (integrated product of voltage and current) without accounting for the signal lag, the blue curve shows the energy calculation after accounting for this lag. The remaining lines in gray, spaced at 200ps increments of $t_{lag}$ indicate the sensitivity to this parameter.
Before each sequence of tests, several hundred pure capacitive pulses were measured and an average lag was determined. The standard deviation of the lag calculated in 1000 calibration cases was 117ps. Since the scope resolution is discretized to 100ps, this indicates that the calculated signal lag was repeatable to within about $\pm$2 discrete shifts. Based on figure~\ref{fig:lag_demo} the error in energy measurement caused by variation in the lag measurement is less than $\pm20\mu J$. This is expected to be the dominant source of systematic error and is on the order of the random error in energy measurement seen between tests (see, for example, figure~\ref{fig:flame_pos_norm}). Overall, the energy measurements are expected to have an uncertainty on the order of tens of $\mu J$.\\
\begin{figure}
    \centering
    \includegraphics[width=0.45\textwidth,trim={1cm 0.5cm 2cm 1.5cm},clip]{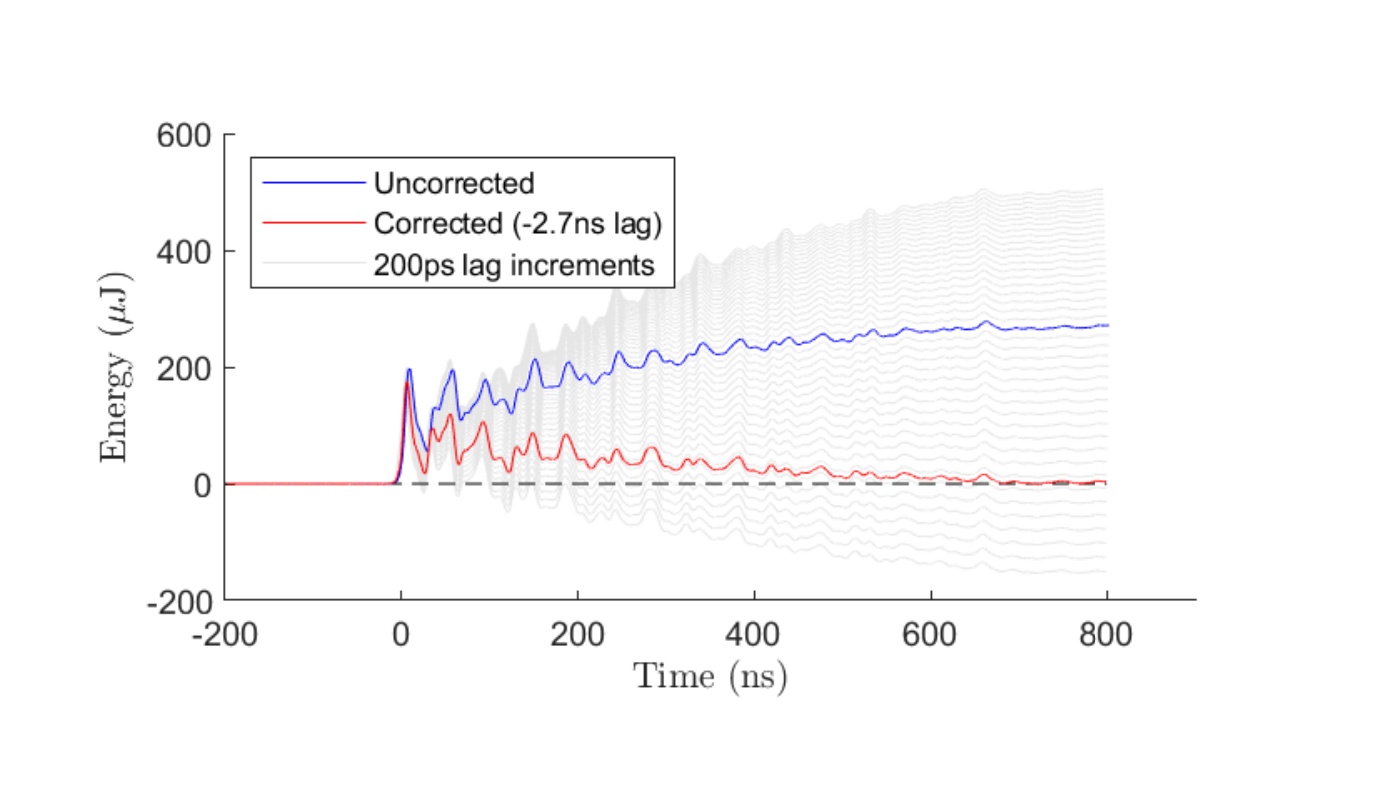}
    \caption{Sensitivity of energy calculation to synchronization of signals. Blue curve is without considering lag, red curve is after calibration, gray curves demonstrate the sensitivity in the calculation to 200ps shifts in lag. \adda{Conditions: f=4kHz, V=8kV, no flow, N$_p$=1600.}}
    \label{fig:lag_demo}
\end{figure}
An alternate method of determining the lag was also considered. For this method, the current and time derivative of the voltage signals in a pure capacitive case were Fourier decomposed and reconstructed using only the common dominant frequencies. The cross-correlation of the reconstructed signals was then used to determine the lag. This method allowed removing the effect of high frequency reflections in the current signal, which represent a real current across the load and must be considered when calculating the energy, but make synchronization of the signals based on a pure-capacitor assumption difficult. The lag calculated by this method was similar to that calculated by assuming the pure capacitor dissipated zero energy, but was more computationally intensive and found to have greater variance than the other method, hence why the zero-energy condition was used.
\\
\section{Discharge Re-ignition Driven by Invigoration of Flame}\label{sec:app_reinvig}
It was noted in the authors' previous work \cite{Pavan2021a} that application of the \dela{RPN}\adda{NRP} DBD discharge could cause premature flame quenching, with the effect more significant at higher pulse repetition frequencies. The electrode position for this experiment resulted in an interesting behaviour when the pulse repetition frequency was set such that the flame would almost quench at the edge of the electrode. In these cases, as the flame travelled along the electrodes, the flame would become weaker and weaker and, consequently, the discharge would almost completely disappear due to a lack of the hot gases needed to increase the reduced electric field to the point that a discharge would occur. When this weak flame did exit the electrode, it would then become stronger again and resume production of large amounts of hot gas. As this gas is convected along the channel behind the flame, it re-ignites a uniform discharge under the electrode. This hot gas is propagating in the same direction of the unburnt gas flow, and so in these cases the uniform discharge will grow from the narrow to the wide end of the channel. This is shown visually in figure~\ref{fig:discharge_reignition}, and agrees with the energy measurements shown in  figure~\ref{fig:long_elec_nrg_reinvig}. At 274ms, the flame has almost extinguished and at 280-284ms there is not enough hot gas being produced to sustain the discharge. By 287ms, as the flame exits the electrodes, the flame front has begun to revive (as evidenced by the higher luminosity) and the hot reaction products begin to reignite the uniform discharge at the narrow end of the inter-electrode region. The speed at which this uniform discharge region grows is approximately 1m/s; faster than the laminar flame speed, but comparable with the speed at which the burnt gas behind the flame front will be moving. This indicates that the growing discharge is caused by the renewed production of hot combustion products by the flame front propagating to the right.
\begin{figure}
    \centering
    \includegraphics[width=0.48\textwidth, trim={1cm 0 1cm 0},clip]{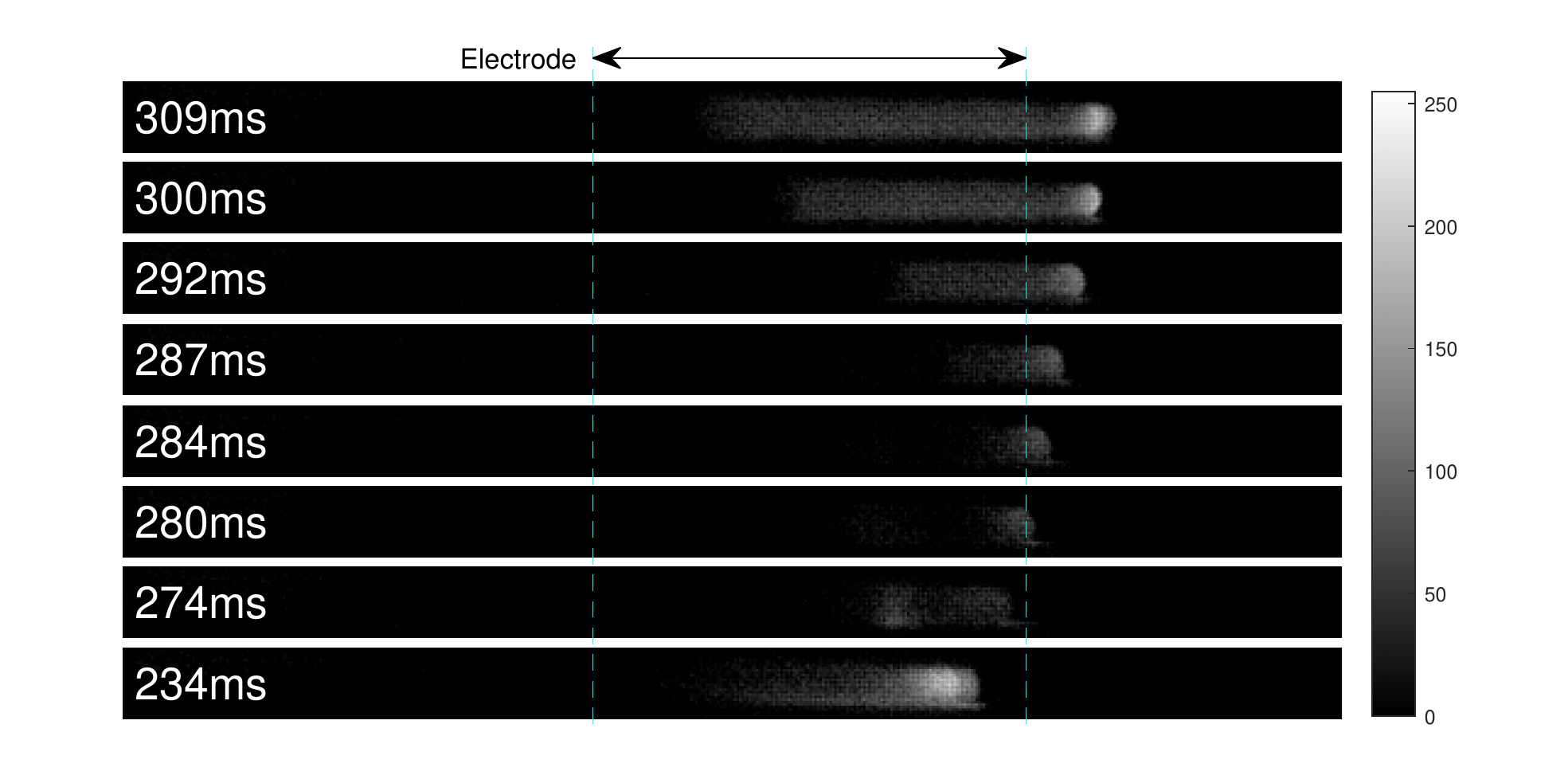}
    \caption{Re-ignition of the uniform discharge caused by a revival of the flame post electrode. Case of \adda{f=6kHz,  V=11.7kV,$\mathrm{\phi}$=1, Q=1000sccm, N$_p$=2400. Legend in colorbar is in arbitrary units.}}
    \label{fig:discharge_reignition}
\end{figure}
\begin{figure}[]
    \centering
    \includegraphics[width=0.45\textwidth, trim={1.3cm 0 1.5cm 0},clip]{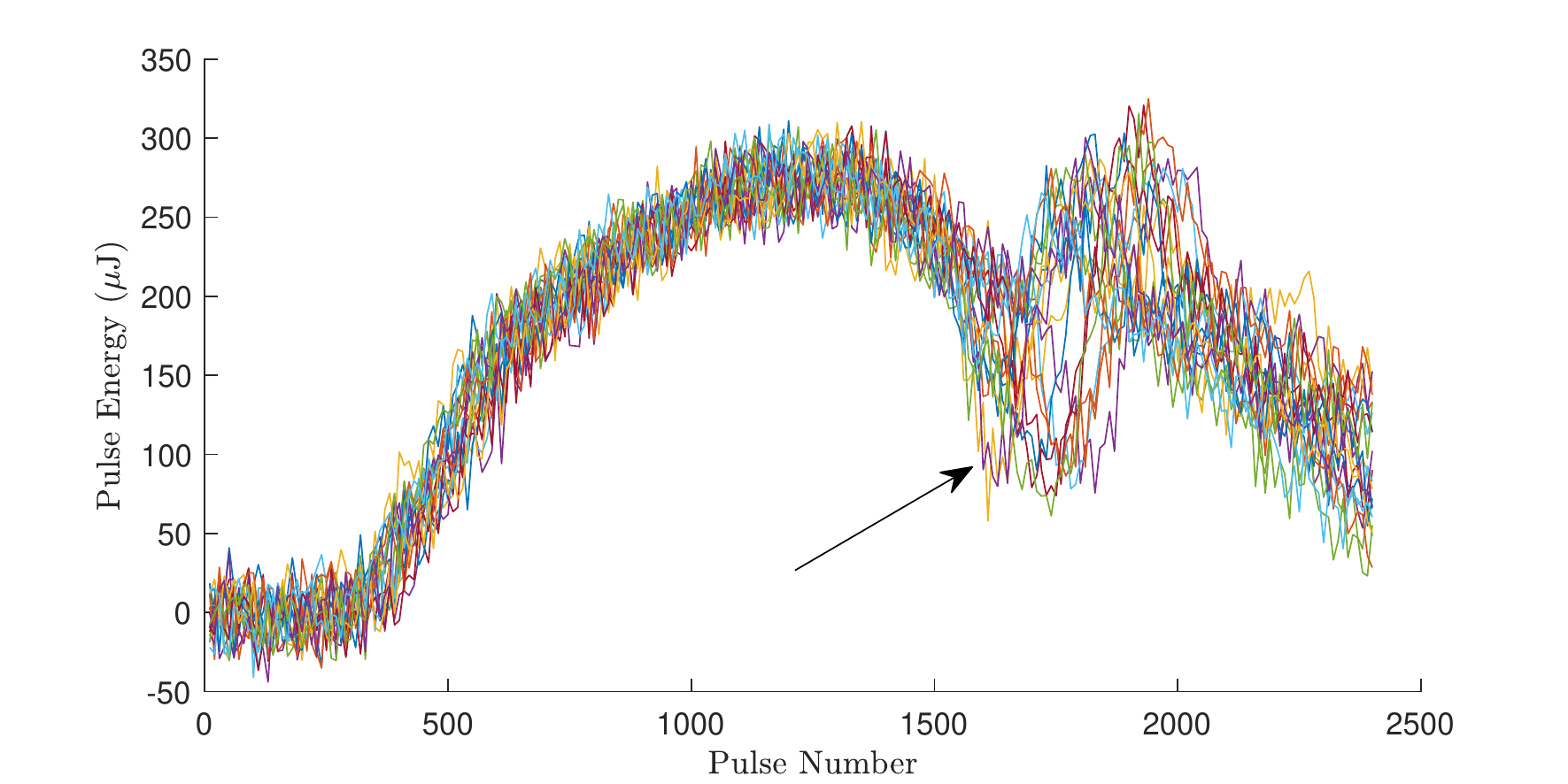}
    \caption{Evolution of the per-pulse energy in a case with partial flame quenching. The arrow indicates the energy decrease when the flame is almost extinguished.  \adda{f=6kHz, V=11.7kV,$\mathrm{\phi}$=1, Q=1000sccm, N$_p$=2400.}}
    \label{fig:long_elec_nrg_reinvig}
\end{figure}
}
\section*{References}
\bibliographystyle{iopart-num}
\bibliography{references.bib}

\providecommand{\newblock}{}
\begin{thebibliography}{10}
\expandafter\ifx\csname url\endcsname\relax
  \def\url#1{{\tt #1}}\fi
\expandafter\ifx\csname urlprefix\endcsname\relax\def\urlprefix{URL }\fi
\providecommand{\eprint}[2][]{\url{#2}}

\bibitem{Lacoste2013}
Lacoste D~A, Xu D~A, Moeck J~P and Laux C~O 2013 {\em Proceedings of the
  Combustion Institute\/} {\bf 34} 3259--3266

\bibitem{Blanchard2021}
Blanchard V~P, Minesi N, Stepanyan S, Stancu G~D and Laux C~O 2021 {Dynamics of
  a Lean Flame Stabilized by Nanosecond Discharges} {\em AIAA Scitech 2021
  Forum\/} January (Reston, Virginia: American Institute of Aeronautics and
  Astronautics) pp 1--10 ISBN 978-1-62410-609-5
  \urlprefix\url{https://arc.aiaa.org/doi/10.2514/6.2021-1700}

\bibitem{Starikovskaia2006a}
Starikovskaia S~M 2006 {\em Journal of Physics D: Applied Physics\/} {\bf 39}
  R265--R299

\bibitem{Pilla2006}
Pilla G, Galley D, Lacoste D~A, Lacas F, Veynante D and Laux C~O 2006 {\em IEEE
  Transactions on Plasma Science\/} {\bf 34} 2471--2477

\bibitem{Barbosa2015}
Barbosa S, Pilla G, Lacoste D~A, Scouflaire P, Ducruix S, Laux C~O and Veynante
  D 2015 {\em Philosophical Transactions of the Royal Society A: Mathematical,
  Physical and Engineering Sciences\/} {\bf 373}

\bibitem{Ju2015}
Ju Y and Sun W 2015 {\em Progress in Energy and Combustion Science\/} {\bf 48}
  21--83

\bibitem{Starikovskiy2013}
Starikovskiy A and Aleksandrov N 2013 {\em Progress in Energy and Combustion
  Science\/} {\bf 39} 61--110

\bibitem{Do2010}
Do H, Cappelli M~A and Mungal M~G 2010 {\em Combustion and Flame\/} {\bf 157}
  1783--1794

\bibitem{Do2010a}
Do H, Im S~K, Cappelli M~A and Mungal M~G 2010 {\em Combustion and Flame\/}
  {\bf 157} 2298--2305

\bibitem{Aleksandrov2009}
Aleksandrov N~L, Kindysheva S~V, Kosarev I~N, Starikovskaia S~M and
  Starikovskii A~Y 2009 {\em Proceedings of the Combustion Institute\/} {\bf
  32} 205--212

\bibitem{Castela2016}
Castela M, Fiorina B, Coussement A, Gicquel O, Darabiha N and Laux C~O 2016
  {\em Combustion and Flame\/} {\bf 166} 133--147

\bibitem{Cathey2007}
Cathey C~D, Tang T, Shiraishi T, Urushihara U, Kuthi A and Gundersen M~A 2007
  {\em IEEE Transactions on Plasma Science\/} {\bf 35} 1664--1668

\bibitem{Kosarev2008}
Kosarev I~N, Aleksandrov N~L, Kindysheva S~V, Starikovskaia S~M and
  Starikovskii A~Y 2008 {\em Combustion and Flame\/} {\bf 154} 569--586

\bibitem{Sun2014}
Sun W, Won S~H and Ju Y 2014 {\em Combustion and Flame\/} {\bf 161} 2054--2063

\bibitem{Yang2017}
Yang S, Nagaraja S, Sun W and Yang V 2017 {\em Journal of Physics D: Applied
  Physics\/} {\bf 50} 433001

\bibitem{Nagaraja2013}
Nagaraja S, Yang V and Adamovich I 2013 {\em Journal of Physics D: Applied
  Physics\/} {\bf 46}

\bibitem{Popov2016}
Popov N~A 2016 {\em Plasma Sources Science and Technology\/} {\bf 25} 043002

\bibitem{Starikovskiy2015}
Starikovskiy A 2015 {\em Philosophical Transactions of the Royal Society A:
  Mathematical, Physical and Engineering Sciences\/} {\bf 373} 20150074

\bibitem{Orr2020}
Orr K, Yang X, Gulko I and Adamovich I~V 2020 {\em Plasma Sources Science and
  Technology\/} {\bf 29} 125022

\bibitem{Rousso2020}
Rousso A~C, Goldberg B~M, Chen T~Y, Wu S, Dogariu A, Miles R~B, Kolemen E and
  Ju Y 2020 {\em Plasma Sources Science and Technology\/} {\bf 29} 105012

\bibitem{Stange2005}
Stange S, Kim Y, Ferreri V, Rosocha L~A and Coates D~M 2005 {\em IEEE
  Transactions on Plasma Science\/} {\bf 33} 316--317

\bibitem{Elkholy2018}
Elkholy A, Shoshyn Y, Nijdam S, van Oijen J~A, van Veldhuizen E~M, Ebert U and
  de~Goey L~P 2018 {\em Experimental Thermal and Fluid Science\/} {\bf 95}
  18--26

\bibitem{Evans2017}
Evans M~D, Bergthorson J~M and Coulombe S 2017 {\em Journal of Applied
  Physics\/} {\bf 122}

\bibitem{Adamovich2009}
Adamovich I~V, Nishihara M, Choi I, Uddi M and Lempert W~R 2009 {\em Physics of
  Plasmas\/} {\bf 16} 113505

\bibitem{Zhong2019}
Zhong H, Shneider M~N, Mokrov M~S and Ju Y 2019 {\em Journal of Physics D:
  Applied Physics\/} {\bf 52} 484001

\bibitem{Guerra-Garcia2014}
Guerra-Garcia C and Martinez-Sanchez M 2013 {\em Journal of Physics D: Applied
  Physics\/} {\bf 46} 345204

\bibitem{Guerra-Garcia2015}
Guerra-Garcia C, Martinez-Sanchez M, Miles R~B and Starikovskiy A 2015 {\em
  Plasma Sources Science and Technology\/} {\bf 24} 055010

\bibitem{Li2013}
Li T, Adamovich I~V and Sutton J~A 2013 {\em Combustion Science and
  Technology\/} {\bf 185} 990--998

\bibitem{Nagaraja2015}
Nagaraja S, Li T, Sutton J~A, Adamovich I~V and Yang V 2015 {\em Proceedings of
  the Combustion Institute\/} {\bf 35} 3471--3478

\bibitem{Massa2017}
Massa L and Freund J~B 2017 {\em Combustion and Flame\/} {\bf 184} 208--232

\bibitem{Adamovich2009a}
Adamovich I~V, Choi I, Jiang N, Kim J~H, Keshav S, Lempert W~R, Mintusov E,
  Nishihara M, Samimy M and Uddi M 2009 {\em Plasma Sources Science and
  Technology\/} {\bf 18} 034018

\bibitem{Nikandrov2008}
Nikandrov D~S, Tsendin L~D, Kolobov V~I and Arslanbekov R~R 2008 {\em IEEE
  Transactions on Plasma Science\/} {\bf 36} 131--139

\bibitem{Lou2007}
Lou G, Bao A, Nishihara M, Keshav S, Utkin Y~G, Rich J~W, Lempert W~R and
  Adamovich I~V 2007 {\em Proceedings of the Combustion Institute\/} {\bf 31}
  3327--3334

\bibitem{Liu2018}
Liu C, Fridman A and Dobrynin D 2018 {\em Plasma Research Express\/} {\bf 1}
  015007

\bibitem{Liu2019}
Liu C, Fridman A and Dobrynin D 2019 {\em Journal of Physics D: Applied
  Physics\/} {\bf 52}

\bibitem{Zhong2021}
Zhong H, Shneider M~N and Ju Y 2021 {Stability Analysis of Thermal-Chemical
  Instability in a Weakly Ionized Plasma} {\em AIAA Scitech 2021 Forum\/}
  January (Reston, Virginia: American Institute of Aeronautics and
  Astronautics) pp 1--7 ISBN 978-1-62410-609-5
  \urlprefix\url{https://arc.aiaa.org/doi/10.2514/6.2021-1702}

\bibitem{Khomich2016}
Khomich V~Y, Malanichev V~E, Malashin M~V and Moshkunov S~I 2016 {\em IEEE
  Transactions on Plasma Science\/} {\bf 44} 1349--1352

\bibitem{Zhang2021}
Zhang Y and Guerra-Garcia C 2021 {Discharge Regimes and Transitions in Pulsed
  Nanosecond Dielectric Barrier Discharges at Atmospheric Pressure} {\em AIAA
  AVIATION 2021 FORUM\/} (Reston, Virginia: American Institute of Aeronautics
  and Astronautics) ISBN 978-1-62410-610-1
  \urlprefix\url{https://arc.aiaa.org/doi/10.2514/6.2021-3118}

\bibitem{Huang2014}
Huang B~D, Takashima K, Zhu X~M and Pu Y~K 2014 {\em Journal of Physics D:
  Applied Physics\/} {\bf 47}

\bibitem{Huang2018}
Huang B~D, Carbone E, Takashima K, Zhu X~M, Czarnetzki U and Pu Y~K 2018 {\em
  Journal of Physics D: Applied Physics\/} {\bf 51}

\bibitem{Huang2020}
Huang B, Zhang C, Adamovich I, Akishev Y and Shao T 2020 {\em Plasma Sources
  Science and Technology\/} {\bf 29}

\bibitem{Moeck2013}
Moeck J, Lacoste D, Laux C and Paschereit C 2013 {Control of combustion
  dynamics in a swirl-stabilized combustor with nanosecond repetitively pulsed
  discharges} {\em 51st AIAA Aerospace Sciences Meeting including the New
  Horizons Forum and Aerospace Exposition\/} January (Reston, Virigina:
  American Institute of Aeronautics and Astronautics) pp 1--11 ISBN
  978-1-62410-181-6 \urlprefix\url{https://arc.aiaa.org/doi/10.2514/6.2013-565}

\bibitem{Kim2015}
Kim W, Snyder J and Cohen J 2015 {\em Proceedings of the Combustion
  Institute\/} {\bf 35} 3479--3486

\bibitem{Zhu2017}
Zhu Y, Anand V, Jodele J, Knight E, Gutmark E~J and Burnette D 2017
  Plasma-assisted rotating detonation combustor operation {\em 53rd
  AIAA/SAE/ASEE Joint Propulsion Conference\/} p 4742

\bibitem{Shanbhogue2022}
Shanbhogue S~J, Weibel D~E, del Campo F~G, Guerra-Garcia C and Ghoniem A~F 2022
  {Active Control of Large Amplitude Combustion Oscillations using Nanosecond
  Repetitively Pulsed Plasmas} {\em AIAA Scitech 2022 Forum\/} (San Diego, CA)

\bibitem{Murphy2014}
Murphy D~C, S{\'{a}}nchez-Sanz M and Fernandez-Pello C 2014 {\em Journal of
  Physics: Conference Series\/} {\bf 557} 012076

\bibitem{Berlad1957}
Berlad A~L and Potter A~E~J 1957 {Effect of Channel Geometry on the Quenching
  of Laminar Flames} Tech. Rep. April National Advisory Committee for
  Aeronautics (NACA)

\bibitem{Pancheshnyi2006}
Pancheshnyi S~V, Lacoste D~A, Bourdon A and Laux C~O 2006 {\em IEEE
  Transactions on Plasma Science\/} {\bf 34} 2478--2487

\bibitem{Rusterholtz2013}
Rusterholtz D~L, Lacoste D~A, Stancu G~D, Pai D~Z and Laux C~O 2013 {\em
  Journal of Physics D: Applied Physics\/} {\bf 46}

\bibitem{Laux2003}
Laux C~O, Spence T~G, Kruger C~H and Zare R~N 2003 {\em Plasma Sources Science
  and Technology\/} {\bf 12} 125--138

\bibitem{Saffman1958}
Saffman P~G and Taylor G~I 1958 {\em Proceedings of the Royal Society of
  London. Series A. Mathematical and Physical Sciences\/} {\bf 245} 312--329

\bibitem{Murphy2015}
Murphy D~C 2015 {\em The Measurement and Application of Electric Effects in
  Combustion\/} Ph.D. thesis UC Berkeley
  \urlprefix\url{https://escholarship.org/uc/item/47s6w7dm}

\bibitem{Lewis1987}
Lewis B and von Elbe G 1987 {\em {Combustion, Flames and Explosions of
  Gases}\/} 3rd ed (Orlando, FL: Academic Press, Inc.) ISBN 0-12-446751-2

\bibitem{Ju2006}
Ju Y and Xu B 2006 {\em Combustion Science and Technology\/} {\bf 178}
  1723--1753

\bibitem{Lefkowitz2015}
Lefkowitz J~K, Uddi M, Windom B~C, Lou G and Ju Y 2015 {\em Proceedings of the
  Combustion Institute\/} {\bf 35} 3505--3512

\bibitem{Pavan2021a}
Pavan C~A and Guerra-Garcia C 2021 {Plasma Actuation of Mesoscale Flames} {\em
  AIAA AVIATION 2021 FORUM\/} (Reston, Virginia: American Institute of
  Aeronautics and Astronautics) ISBN 978-1-62410-610-1
  \urlprefix\url{https://arc.aiaa.org/doi/10.2514/6.2021-3105}

\bibitem{cantera}
Goodwin D~G, Speth R~L, Moffat H~K and Weber B~W 2018 {Cantera: An
  Object-oriented Software Toolkit for Chemical Kinetics, Thermodynamics, and
  Transport Processes} \urlprefix\url{https://cantera.org/}

\bibitem{Bergman2011}
Bergman T, Lavine A, Incropera F and Dewitt D 2011 {\em {Fundamentals of Heat
  and Mass Transfer}\/} seventh ed (John Wiley {\&} Sons, Inc) ISBN
  978-0470-50197-9

\bibitem{Pavan2022}
Pavan C~A and Guerra-Garcia C 2022 Modelling the impact of a repetitively
  pulsed nanosecond dbd plasma on a mesoscale flame {\em AIAA SCITECH 2022
  Forum\/} p 0975

\bibitem{Liu2014}
Liu C, Dobrynin D and Fridman A 2014 {\em Journal of Physics D: Applied
  Physics\/} {\bf 47}

\bibitem{Fridman2011}
Fridman A~A and Kennedy L~A 2011 {\em {Plasma physics and engineering}\/} 2nd
  ed (Boca Raton, FL: Taylor \& Francis Group) ISBN 978-1-4398-1228-0

\bibitem{Raizer1991}
Raizer Y~P 1991 {\em {Gas Discharge Physics}\/} (Berlin: Springer-Verlag Berlin
  Heidelberg) ISBN 0-387-19462-2

\bibitem{Gadkari2017}
Gadkari S and Gu S 2017 {\em Physics of Plasmas\/} {\bf 24}

\end{thebibliography}

\end{document}